\documentclass[twocolumn,aps,prb,showpacs,longbibliography,10pt]{revtex4-2} 

\usepackage[normalem]{ulem}
\usepackage{amsfonts,amssymb,amsmath,graphicx,color}
\usepackage{array,multirow}
\usepackage{mathtools}
\usepackage{enumitem}
\usepackage{braket}
\usepackage{physics}
\usepackage[version=4]{mhchem}
\usepackage{float}
\usepackage[colorlinks=true,linkcolor=blue,citecolor=blue,urlcolor=blue]{hyperref}

\newcommand\commentout[1]{}

\newcolumntype{P}[1]{>{\centering\arraybackslash}p{#1}}

\definecolor{greenPR}{rgb}{0.00, 0.6, 0.00}

\begin{document}
	
	\title{Designing flat-band tight-binding models with tunable multifold band touching points}
	\author{Ansgar Graf}
	\author{Fr\'ed\'eric Pi\'echon}
	\affiliation{%
		Universit\'e Paris-Saclay, CNRS, Laboratoire de Physique des Solides, 91405, Orsay, France\\
	}%
	\date{\today}
	
	\begin{abstract}
Being dispersionless, flat bands on periodic lattices are solely
characterized by their macroscopically degenerate
eigenstates: compact localized
states (CLSs) in real space and Bloch states in reciprocal space.
Based on this property, this work presents a straightforward method to build flat-band tight-binding models with short-range hoppings \emph{on any periodic lattice}. The method consists in starting from a CLS and engineering families of Bloch Hamiltonians as quadratic (or linear) functions
of the associated Bloch state. The resulting tight-binding models not only exhibit a flat band, but also multifold quadratic (or linear)
band touching points
(BTPs) whose number, location, degeneracy and (non-)singularity can be controlled to a large extent. Quadratic flat-band models are ubiquitous: they can be built from any arbitrary CLS, on any lattice, in any dimension and with any number $N\geq2$ of
bands. Linear flat-band models are rarer: they require $N\geq3$ and can only be built from CLSs that fulfill certain compatibility relations with the underlying lattice. Most flat-band models from the literature can be classified according to this scheme: Mielke's and Tasaki's models belong to
the quadratic class, while the Lieb, dice and breathing Kagome models belong
to the linear class.
Many novel flat-band models are introduced, among which an $N=4$ bilayer honeycomb model with fourfold quadratic BTPs, an $N=5$ dice model with fivefold linear BTPs, and an $N=3$ Kagome model with BTPs that can be smoothly tuned from linear to quadratic.
\end{abstract}
	\maketitle

\section{Introduction}

 The energy band structure associated to quantum mechanical particles on a periodic lattice is a fundamental building stone of solid state physics \cite{Ashcroft_1976}, and the concept can be naturally extended to artificial periodic systems, for example photonic crystals \cite{Joannopoulos_2011}. Currently, particular attention is paid to band structures that exhibit interesting features with regard to quantum geometry, topology and interactions. In this paper, we are interested in two such features, namely (1) flat bands and (2) band touching points.
 
(1) A \emph{flat band} is a completely dispersionless energy band. Flat bands have long been known to develop in a two-dimensional electron gas (2DEG) in the presence of a magnetic field (Landau levels) \cite{Landau_1930}, and were first introduced in the context of lattice models in the seminal works of Sutherland, Lieb, Mielke and Tasaki \cite{Sutherland_1986,Lieb_1989,Mielke_1991,Mielke_1991b,Tasaki_1992}. 
 
Flat-band systems are interesting because, on the one hand, band flatness signifies the absence of an intrinsic energy scale (a band width), and on the other hand it implies the existence of macroscopically degenerate eigenstates. These two properties combined ensure that any perturbation may act non-perturbatively and profoundly modify the physics of the flat band. As an emblematic example, if electron-electron (Coulomb) interactions are added to a Landau-quantized 2DEG, the effect is known to be strongly non-perturbative, and the possibility for the existence of strongly correlated quantum phases arises (fractional quantum Hall states \cite{Girvin_2005}, Wigner crystal \cite{Monarkha_2012}, etc.).


While the first works on flat-band lattices were mostly interested in ferromagnetism induced by the Hubbard interaction \cite{Tasaki_1998,Derzhko_2015}, flat-band lattices are now widely studied in many electronic and artificial systems \cite{Liu_2014,Leykam_2018}. Moreover, almost-flat bands can be found in Kagome-type materials, see for instance Refs. \cite{Lin_2018,Kang_2020,Liu_2020}, as well as in twisted bi- and multilayers of graphene and other quasi-2D crystals \cite{Cheng_2019,Andrei_2020,Mogera_2020}. In the latter class of materials, an astonishing variety of different regimes emerges from the nearly-flat bands, ranging from quantum Hall to Mott insulating and superconducting phases.

The simplest approach to studying flat-band systems consists in considering periodic lattices in the tight-binding (TB) approximation; this is the framework adopted in the present work. In particular, many researchers have tried to develop schemes that systematically generate flat-band TB models. Such schemes make use of graph theory \cite{Mielke_1991,Mielke_1991b,Miyahara_2005,Roentgen_2018,Morfonios_2021}, Origami rules \cite{Dias_2015}, repeated miniarrays \cite{Moral_2016}, a bipartite lattice structure \cite{Ramachandran_2017,Calugaru_2021}, generic existence conditions \cite{Toikka_2018,Ogata_2021}, or the extension of known flat-band lattices \cite{Mizoguchi_2019,Lee_2019,Ogata_2021}. Also, over the last years, the close relation between flat bands and \textit{compact localized states (CLSs)} \cite{Aoki_1996} has been increasingly exploited \cite{Nishino_2003,Nishino_2005,Flach_2014,Maimaiti_2017,Maimaiti_2019,Rhim_2019,Maimaiti_2021,Sathe_2021}. A CLS is a wave function strictly localized to some finite region of the lattice, with zero probability amplitude outside this region.

(2) A \emph{band touching point (BTP)} is a point in the first Brillouin zone (FBZ) where two (twofold BTP) or more (multifold BTP) energy bands touch. Such defects generically entail geometrical and topological structures that have a measurable impact on physical observables. Accordingly, they have become a major focus of condensed matter research and are investigated in all kinds of setups, from graphene \cite{Castro_2009} to Dirac and Weyl semimetals \cite{Armitage_2018} and even non-Hermitian systems \cite{Ashida_2020}. 
Interestingly, the notions of flat bands and BTPs are not entirely independent. 
 Some generic properties of BTPs in flat-band systems have been studied in Refs. \cite{Bergman_2008,Rhim_2019,Mizoguchi_2019b}. 

\begin{figure}
	\centering
	\includegraphics[width=\columnwidth]{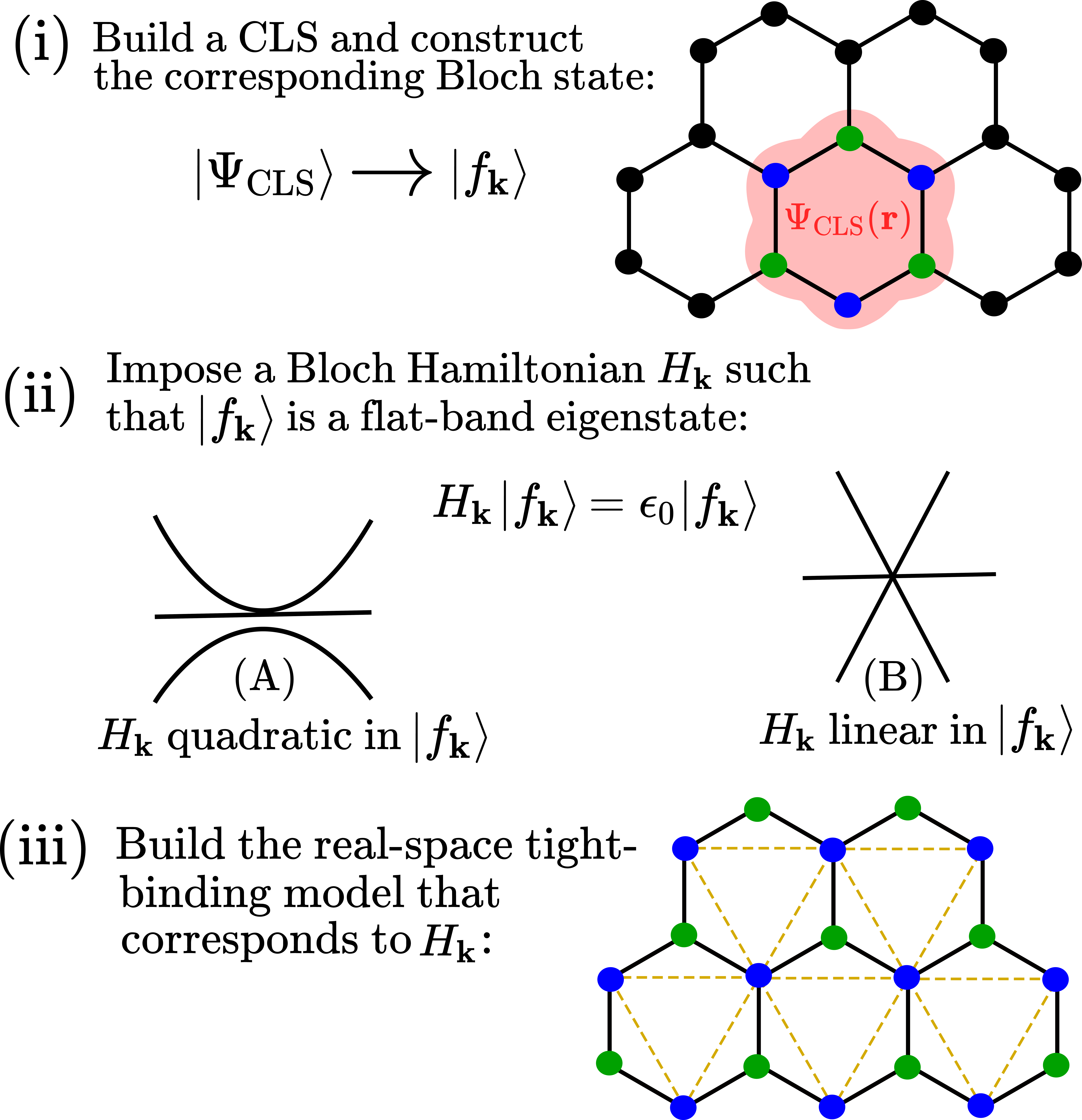}
	\caption{Basics of our method to build flat-band lattice models. In step (i), the CLS wave function in real space is illustrated by the red shading. In step (ii), $\epsilon_0$ denotes the energy of the flat band. An example for a band structure with a quadratic (linear) BTP is sketched on the left (right). In step (iii), a schematic visualization of the final flat-band hopping model is shown.}
	\label{fig:method}
\end{figure}

Here, we are interested in families of tight-binding models that exhibit both features (1) and (2). In particular, we propose a systematic and simple procedure to design flat-band TB models, which at the same time offers considerable control over the existence and character of multifold BTPs.
 The key idea consists in an inverse approach: instead of directly searching for a real-space TB Hamiltonian with a flat band, we \emph{impose} a form of the Bloch Hamiltonian $H_\mathbf{k}$ that forces a flat band to occur in its band structure. An appropriate real-space Hamiltonian can then be built in a second step from the knowledge of $H_\mathbf{k}$.
     This procedure is outlined in more detail in Fig. \ref{fig:method}. 

 (i) On any lattice of choice, build any CLS $\ket{\Psi_\text{CLS}}$, i.e. a linear combination of a finite number of atomic orbitals. The corresponding Bloch state $\ket{f_\mathbf{k}}$ is formed as a superposition of all copies of the CLS translated by a Bravais vector. (ii) Impose a Bloch Hamiltonian $H_\mathbf{k}(\ket{f_\mathbf{k}})$ such that $\ket{f_\mathbf{k}}$ is automatically a flat-band eigenstate. In particular, $H_\mathbf{k}(\ket{f_\mathbf{k}})$ can exhibit a quadratic or linear dependency on $\ket{f_\mathbf{k}}$, generically leading to models with quadratic or linear BTPs, respectively. (iii) Build the real-space TB Hamiltonian corresponding to $H_\mathbf{k}$, which will be a flat-band model by construction. 
 
 This method has several very convenient features. First, it allows to obtain arbitrarily many flat-band tight-binding models \emph{on any periodic lattice}. Second, it is conceptually straightforward and fully analytic. Third, the required hoppings are generically short-ranged and do not require fine-tuning. Moreover, since $H_\mathbf{k}$ is obtained before the real-space model, the band structure can be directly controlled without the explicit need to construct the real-space model first. In particular, the flat band can be easily designed to be singular or non-singular \cite{Rhim_2019}, the dispersive bands can be widely tuned without destroying the flat band, and multifold singular or non-singular BTPs with a tunable degree of degeneracy can be created. Finally, the method can be viewed as a new classification scheme, placing known models (Mielke, Tasaki, Lieb, dice, breathing Kagome, etc.) and the models we propose into quadratic or linear classes, based on the relation between the Bloch Hamiltonian and the CLS. 
 
 This paper is organised as follows. In Section \ref{latticesec} we first recall the important notion of a CLS and how to construct the corresponding Bloch state $\ket{f_\mathbf{k}}$ (Section \ref{secCLS}). Second, the key idea of this work is introduced (Section \ref{Blochconstr}): flat-band Bloch Hamiltonians can be \emph{engineered} from any given CLS. This idea is developed for the case of a quadratic relation between $H_\mathbf{k}$ and $\ket{f_\mathbf{k}}$ in Section \ref{quadmods}, and for the case of a linear relation in Section \ref{linmods}. CLSs that allow to build linear flat-band models can also be used to build models that smoothly interpolate between the linear and quadratic case, as discussed in Section \ref{mixedmods}. Conclusions are given in Section \ref{concl}.

\section{Building flat-band Hamiltonians from compact localized states}
\label{latticesec}

\subsection{Compact localized states and associated Bloch states}
\label{secCLS}

 Consider a periodic lattice treated in the TB approximation \cite{Ashcroft_1976}, with $N$ (atomic) orbitals per unit cell, i.e. with $N$ sublattices labeled by $\alpha=A,B,C,...$\,. On such a lattice, a \emph{compact localized state (CLS)} centered at some localization center $\mathbf{R}_\text{C}$ can be formed as a linear combination 
 \begin{equation}
 	|\Psi_\text{CLS}^{\mathbf{R}_\text{C}}\rangle=\sum_{\alpha_i\in\text{CLS}}w_{\alpha_i}\ket{\alpha_i}
 	\label{CLSdef}
 \end{equation}
 of orbitals $\ket{\alpha_i}$, where the complex number $w_{\alpha_i}$ -- the \emph{CLS amplitude} on the respective orbital -- takes a nonzero value only for a finite number of lattice sites around $\mathbf{R}_\text{C}$. The subindex $i$ takes account of the fact that more than one orbital of a given type $\alpha$ will in general appear in the CLS (orbitals $\alpha_i$ and $\alpha_{j\neq i}$ are in different unit cells). Since only a finite number of orbitals contributes to the CLS, the CLS wave function
 \begin{equation}
 \Psi_\text{CLS}^{\mathbf{R}_\text{C}}(\mathbf{r})\propto\langle\mathbf{r}|\Psi_\text{CLS}^{\mathbf{R}_\text{C}}\rangle
 \end{equation}
is restricted to a finite region of the lattice, with strictly zero probability amplitude outside this region. 
 
\begin{figure}
 	\centering
 	\includegraphics[width=\columnwidth]{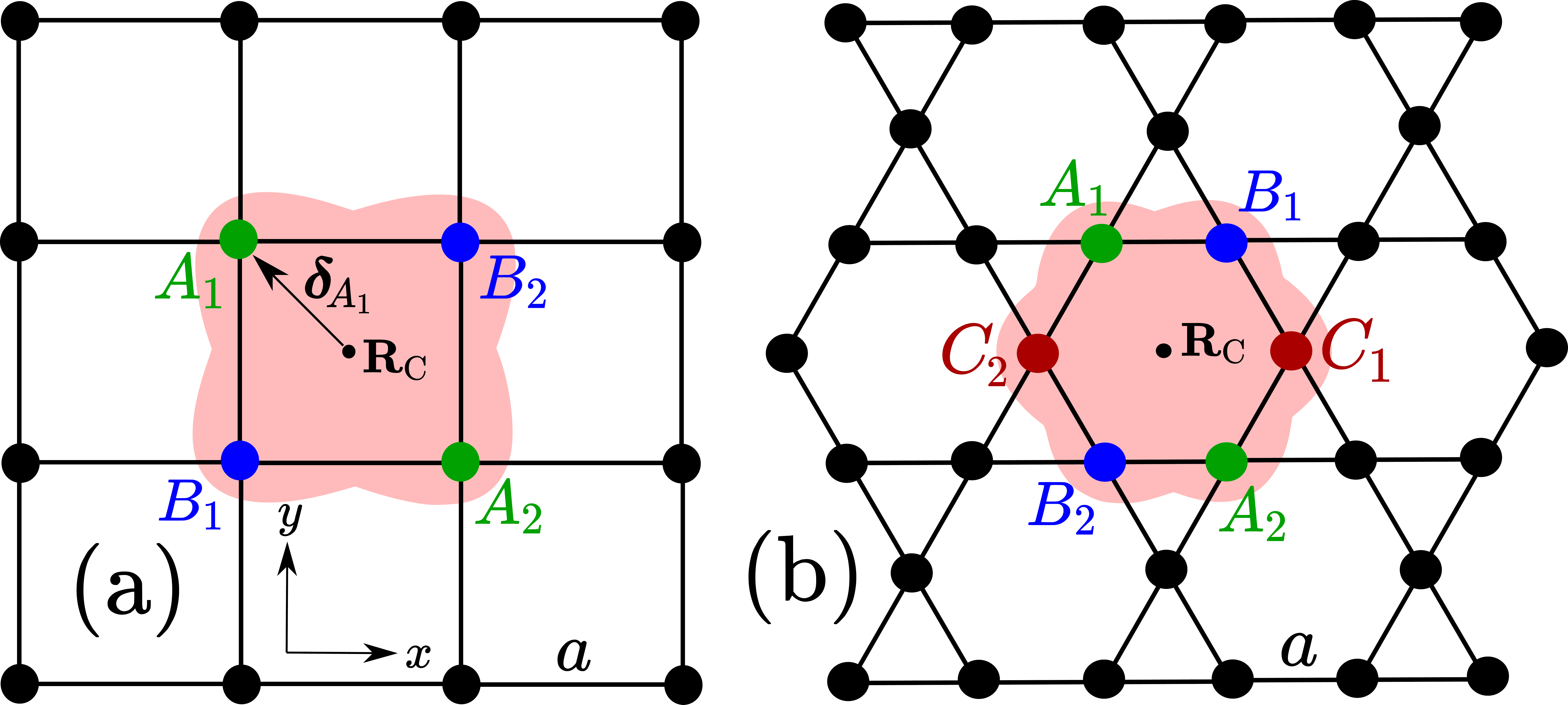}
 	\caption{Example for a CLS (a) on a square lattice ($N=2$) and (b) on a Kagome lattice ($N=3$). The orbitals included in the CLS are colored and labeled. The CLS wave function is represented by the light red region.}
 	\label{fig:newCLSexample}
 \end{figure}

  As an example, consider the CLS built on the square lattice ($N=2$) shown in Fig. \ref{fig:newCLSexample}(a). 
 It involves four orbitals labeled by $\alpha=A,B$ and $i=1,2$, i.e.
 \begin{equation}
 	\begin{aligned}
|\Psi_\text{CLS}^{\mathbf{R}_\text{C}}\rangle&=w_{A_1}\ket{A_1}+w_{A_2}\ket{A_2}\\
&+w_{B_1}\ket{B_1}+w_{B_2}\ket{B_2}.
\end{aligned}
 \end{equation}
  Similarly, the CLS on the Kagome lattice ($N=3$) shown in Fig. \ref{fig:newCLSexample}(b) involves six orbitals, i.e.
 \begin{equation}
	\begin{aligned}
		|\Psi_\text{CLS}^{\mathbf{R}_\text{C}}\rangle&=w_{A_1}\ket{A_1}+w_{A_2}\ket{A_2}\\
		&+w_{B_1}\ket{B_1}+w_{B_2}\ket{B_2}\\
		&+w_{C_1}\ket{C_1}+w_{C_2}\ket{C_2}.
	\end{aligned}
\end{equation}
The orbitals with nonzero amplitude ($w_{\alpha_i}\neq0$) are located at positions $\smash{\boldsymbol{\delta}_{\alpha_i}}$ as measured from $\mathbf{R}_\text{C}$. For the CLS of panel (a), we have $\smash{\boldsymbol{\delta}_{A_{1,2}}=\mp\frac{1}{2}}(1,-1)$ and $\smash{\boldsymbol{\delta}_{B_{1,2}}=\mp\frac{1}{2}}(1,1)$. For the CLS of panel (b), the positions are $\smash{\boldsymbol{\delta}_{A_{1,2}}=\mp\frac{1}{2}}(1,-\sqrt{3})$, $\smash{\boldsymbol{\delta}_{B_{1,2}}=\pm\frac{1}{2}}(1,\sqrt{3})$, and $\smash{\boldsymbol{\delta}_{C_{1,2}}=\pm(1,0)}$. Note that the lattice constant is taken as $a=1$ here and throughout.
 
For any CLS (\ref{CLSdef}) centered at some $\mathbf{R}_\text{C}$, there exists a macroscopic number of equivalent copies translated by some Bravais vector. Thus, for any given CLS, we can introduce a \emph{Bloch state of the CLS (BCLS)} $\ket{f_\mathbf{k}}$, which is obtained as an unnormalized superposition of the CLS and all its translated copies:
\begin{equation}
	\begin{aligned}
	\ket{f_\mathbf{k}}&=\sum_{\mathbf{R}_\text{C}}e^{-i\mathbf{k}\cdot\mathbf{R}_\text{C}}|\Psi_\text{CLS}^{\mathbf{R}_\text{C}}\rangle\\
	&=\sum_\alpha f_\alpha(\mathbf{k}) \ket{\alpha,\mathbf{k}}.
	\end{aligned}
\label{BCLSdef}
\end{equation}
Note that the (reciprocal-space) BCLS is simply the Fourier transform of the (real-space) CLS.
In the second line of Eq. (\ref{BCLSdef}), $\ket{\alpha,\mathbf{k}}$ are Bloch basis states \footnote{The Bloch basis states used here may be written as \smash{$\ket{\alpha,\mathbf{k}}\equiv\mathcal{N}^{-1/2}\sum_me^{-i\mathbf{k}\cdot(\mathbf{R}_m+\mathbf{r}_\alpha)}\ket{m,\alpha}$}. The sum $\sum_m$ runs over all lattice vectors, $\mathcal{N}$ is the total number of unit cells and \smash{$\mathbf{R}_m+\mathbf{r}_\alpha$} is the position vector of the orbital $\ket{m,\alpha}$, i.e. the orbital of type $\alpha$ in unit cell $m$. Note that the BCLS was introduced in a similar way in Ref. \cite{Rhim_2019}, where the authors work in the ''tight-binding basis I'' such that the Bloch Hamiltonian has the periodicity of the FBZ. In contrast, here we work in the ''tight-binding basis II'' \cite{Bena_2009}.} and
 \begin{equation}
	f_\alpha(\mathbf{k})\equiv\sum_{i\in\text{CLS}}w_{\alpha_i}e^{i\mathbf{k}\cdot\boldsymbol{\delta}_{\alpha_i}}
	\label{locdef1}
\end{equation}
is completely specified by the positions and amplitudes of the orbitals contained in the associated real-space CLS. 
Hereafter, we work in the Bloch basis, where the BCLS takes the form of a column vector:
 \begin{equation}
 		\ket{f_\mathbf{k}}=\begin{pmatrix}
 			f_A(\mathbf{k})\\
 			f_B(\mathbf{k})\\
 			...
 		\end{pmatrix}.
 		\label{locdef}
 \end{equation}
For example, the CLS of Fig. \ref{fig:newCLSexample}(a) gives rise to a BCLS
 \begin{equation}
 	\ket{f_\mathbf{k}}=\begin{pmatrix}
 		w_{A_1}e^{-\frac{i}{2}(k_x-k_y)}+w_{A_2}e^{\frac{i}{2}(k_x-k_y)}\\
 		w_{B_1}e^{-\frac{i}{2}(k_x+k_y)}+w_{B_2}e^{\frac{i}{2}(k_x+k_y)}
 	\end{pmatrix},
 	\label{squaref0}
 \end{equation}
and the CLS of Fig. \ref{fig:newCLSexample}(b) corresponds to a BCLS
 \begin{equation}
	\ket{f_\mathbf{k}}=\begin{pmatrix}
		w_{A_1}e^{-ik_-}+w_{A_2}e^{ik_-}\\
		w_{B_1}e^{ik_+}+w_{B_2}e^{-ik_+}\\
		w_{C_1}e^{ik_x}+w_{C_2}e^{-ik_x}
	\end{pmatrix},
	\label{kagomef}
\end{equation}
where $k_\pm\equiv\frac{1}{2}(k_x\pm \sqrt{3}k_y)$. In the same way, each arbitrarily designed CLS on any lattice uniquely corresponds to some BCLS $\ket{f_\mathbf{k}}$; the latter is an important quantity in our flat-band construction scheme, cf. Fig. \ref{fig:method}. 
 
 Note that $\ket{f_\mathbf{k}}$ is called \emph{singular} if there exists a point $\mathbf{k}_0$ in the FBZ such that $\ket{f_{\mathbf{k}_0}}=0$, while it is called \emph{non-singular} otherwise \cite{Rhim_2019}. The (non-)singularity of $\ket{f_\mathbf{k}}$ is a useful tool for characterizing the properties of BTPs at the flat-band energy.
 
 \subsection{Bloch Hamiltonians from compact localized states}
 \label{Blochconstr}
 
We now argue that for any arbitrarily shaped (input) CLS on any periodic lattice, it is possible to construct families of tight-binding Hamiltonians $H$ such that this CLS and all its translated copies constitute a macroscopically degenerate set of eigenstates:
 \begin{equation} 
 	H|\Psi_\text{CLS}^{\mathbf{R}_\text{C}}\rangle=\epsilon_0 |\Psi_\text{CLS}^{\mathbf{R}_\text{C}}\rangle,\hspace{.5cm}\forall\,\mathbf{R}_\text{C}.
\end{equation}
In order to find these tight-binding Hamiltonians $H$, the strategy adopted in this work consists in \emph{designing} Bloch Hamiltonians $H_\mathbf{k}$ ($N\times N$ matrices) for which the BCLS $\ket{f_\mathbf{k}}$ associated to the input CLS represents a flat-band eigenstate:
  \begin{equation}
 	H_\mathbf{k}\ket{f_\mathbf{k}}=\epsilon_0\ket{f_\mathbf{k}}.
 	\label{eqcon}
 \end{equation}
Once such matrices $H_\mathbf{k}$ are found, it is straightforward to obtain the real-space models $H$ from them. 
Hereafter, we will take $\epsilon_0=0$ without loss of generality.

 To see how this Bloch Hamiltonian design can be done, let us write Eq. (\ref{eqcon}) explicitly in matrix form:
 \begin{equation}
 	\begin{bmatrix}
 		H_{\mathbf{k},AA} & H_{\mathbf{k},AB} & ...\\
 		H_{\mathbf{k},BA} & H_{\mathbf{k},BB}& ...\\
 		... & ... & ...
 	\end{bmatrix}\begin{pmatrix}
 		f_A(\mathbf{k})\\ f_B(\mathbf{k})\\...
 	\end{pmatrix}=0.
 	\label{matcon}
 \end{equation}
 Evidently, \emph{a flat band will automatically exist} if the Bloch Hamiltonian is \emph{imposed} to be an appropriate function $H_\mathbf{k}(\ket{f_\mathbf{k}})$ of the BCLS. More concretely, the matrix elements $H_{\mathbf{k},\alpha\beta}$ should depend on the components $f_\alpha(\mathbf{k})$ in such a way that $\sum_\beta H_{\mathbf{k},\alpha\beta}f_\beta(\mathbf{k})=0$ is fulfilled for every $\alpha$, independently of the detailed functional form of the components $f_\alpha(\mathbf{k})$.
 
 Thus, the task at hand consists in finding matrices $H_\mathbf{k}(\ket{f_\mathbf{k}})$ that fulfill Eq. (\ref{matcon}). However, we have to be aware of an additional physical constraint:
 \begin{equation}
 	H_\mathbf{k}\text{ must be a \emph{reasonable} Bloch Hamiltonian.}
 	\label{constr2}
 \end{equation} 
This means that $H_\mathbf{k}(\ket{f_\mathbf{k}})$ has to be chosen such that the real-space TB Hamiltonian $H$ constructed from it makes sense on a lattice. The latter property is not automatically guaranteed for all matrices $H_\mathbf{k}(\ket{f_\mathbf{k}})$ that fulfill Eq. (\ref{matcon}). More concretely, for $H_\mathbf{k}$ to be reasonable, the matrix elements $H_{\mathbf{k},\alpha\beta}$ should be made of finite sums of Bloch phases that are compatible with the underlying lattice geometry (see Appendix \ref{AppA} for more details).
 
As it turns out, there are two different simple choices for matrices that fulfill condition (\ref{matcon}):
 \[
 \begin{aligned}
 	& H_\mathbf{k}\text{ is a \emph{quadratic} function of the }f_\alpha(\mathbf{k})\\
 	&\rightarrow\text{flat-band models with quadratic BTPs, Section \ref{quadmods}.}\\
 	& H_\mathbf{k}\text{ is a \emph{linear} function of the }f_\alpha(\mathbf{k})\\
 	&\rightarrow\text{flat-band models with linear BTPs, Section \ref{linmods}.}
 \end{aligned}
 \]
 Designing quadratic Hamiltonians is possible for $N\geq2$ and, conveniently, condition (\ref{constr2}) is automatically fulfilled for any input CLS. In contrast, designing linear Hamiltonians requires $N\geq3$ and condition (\ref{constr2}) is only fulfilled for special input CLSs. For this reason, the quadratic case may be regarded as simpler and will be treated first.

\section{Flat-band models with quadratic band touching}
\label{quadmods}   

The simplest scenario for a quadratic flat-band model is encountered in two-band ($N=2$) systems (Section \ref{2quad}). The procedure can then straightforwardly be generalized to $N=3$ (Section \ref{3quad}), and finally to any $N$ (Section \ref{Nquad}).
 
\subsection{Two-band models} 
\label{2quad}

Consider some CLS built on a lattice with two orbitals per unit cell:
\begin{equation}
|\Psi_\text{CLS}^{\mathbf{R}_\text{C}}\rangle=\sum_{i\in\text{CLS}}(w_{A_i}\ket{A_i}+w_{B_i}\ket{B_i}).
\label{2CLS}
\end{equation}
The corresponding BCLS has two components, 
\begin{equation}
\ket{f_\mathbf{k}}=(f_A,f_B)^T,
\label{2BCLS}
\end{equation}
where hereafter $f_\alpha\equiv f_\alpha(\mathbf{k})$ for brevity. We now want to construct a matrix $H_\mathbf{k}$ that vanishes on $\ket{f_\mathbf{k}}$ (i.e. $H_\mathbf{k}\ket{f_\mathbf{k}}=0$). The only generic way to do this consists in introducing a state
\begin{equation} \ket{f^{AB}_\mathbf{k}}\equiv(-f_B^*,f_A^*)^T
	\end{equation}
 orthogonal to $\ket{f_\mathbf{k}}$, as well as the matrix
 \begin{equation}
 F^{AB}_\mathbf{k}\equiv\ket{f^{AB}_\mathbf{k}}\bra{f^{AB}_\mathbf{k}}.
 \end{equation}
Then a Bloch Hamiltonian defined as
\begin{equation}
	H_\mathbf{k}\equiv\lambda^{AB}_\mathbf{k}F^{AB}_\mathbf{k}=\lambda^{AB}_\mathbf{k}\begin{bmatrix}
		|f_B|^2 & -f_A f_B^*\\
		-f_A^* f_B & |f_A|^2
	\end{bmatrix}
	\label{2Ham}
\end{equation}
 will obviously have $\ket{f_\mathbf{k}}$ as an eigenstate of zero energy; in other words, Eq. (\ref{matcon}) is fulfilled for any $\ket{f_\mathbf{k}}$.
  The band structure associated to Eq. (\ref{2Ham}) reads
 \begin{equation}
 	\begin{aligned}
 		\epsilon_0(\mathbf{k})&=0,\\
 		\epsilon_1(\mathbf{k})&=\lambda^{AB}_\mathbf{k}(|f_A|^2+|f_B|^2).
 	\end{aligned}
 	\label{2bands}
 \end{equation}
Here, \smash{$\lambda^{AB}_\mathbf{k}$} is an arbitrary real function with the periodicity of the FBZ. If it is taken positive (or negative) throughout, the flat band is gapped away from the dispersive band at all $\mathbf{k}$-points where $\ket{f_\mathbf{k}}$ does not vanish, i.e. only singular BTPs \cite{Rhim_2019} can occur. In contrast, if \smash{$\lambda^{AB}_\mathbf{k}$} changes sign in the FBZ, non-singular BTPs are possible. Since we are predominantly interested in nontrivial BTPs created by the singularity of the flat-band eigenstate, we will take \smash{$\lambda^{AB}_\mathbf{k}=1$} from now on.

\begin{figure}
	\centering
	\includegraphics[width=\columnwidth]{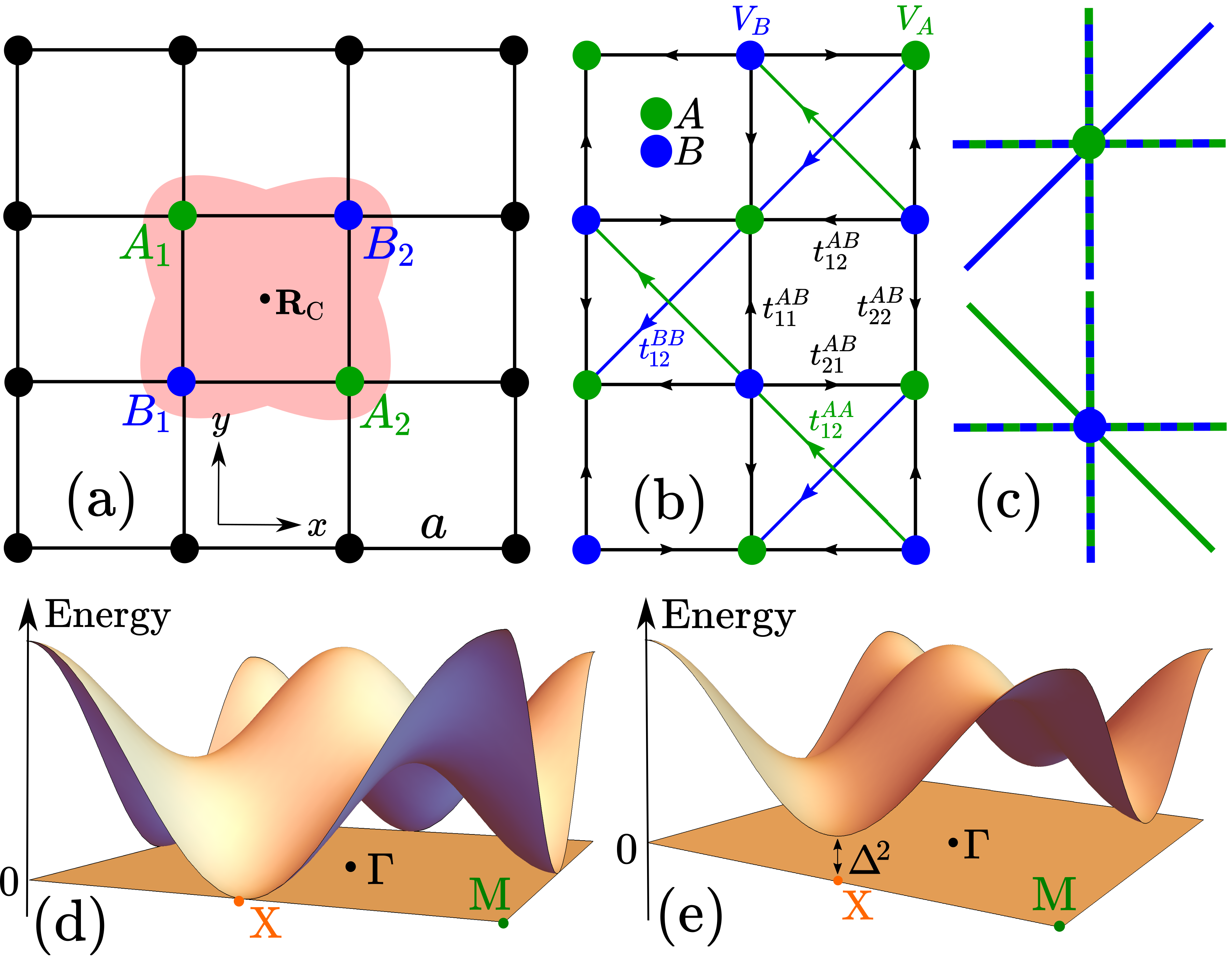}
	\caption{(a) CLS on the square lattice. (b) Flat-band TB model built from it. Hopping parameters are in general complex, as indicated by the arrows. (c) The hopping directions are determined by the CLS shape. (d) Band structure for \smash{$w_{A_i}=-w_{B_i}=1$}. (e) Band structure for the same amplitudes except \smash{$w_{B_2}=-1-\Delta$}.}
	\label{fig:CLSexample}
\end{figure}

Let us now see how flat-band TB models can be designed from Eq. (\ref{2Ham}).  As a first example, consider again the CLS used previously, cf. Fig. \ref{fig:CLSexample}(a).
 We now simply insert the corresponding BCLS (\ref{squaref0}) into the Bloch Hamiltonian (\ref{2Ham}), to obtain
\begin{equation}
H_\mathbf{k}=\begin{bmatrix}
H_{AA} & H_{AB}\\
H_{AB}^* & H_{BB}
\end{bmatrix},
\label{expl2}
\end{equation}
where 
\begin{equation}
\begin{aligned}
H_{AA}&=V_A+t_{12}^{BB}e^{-i(k_x+k_y)}+t_{21}^{BB}e^{i(k_x+k_y)},\\
H_{BB}&= V_B+t_{12}^{AA}e^{-i(k_x-k_y)}+t_{21}^{AA}e^{i(k_x-k_y)},\\
H_{AB}&=t_{11}^{AB}e^{ik_y}+t_{12}^{AB}e^{-ik_x}\\
&+t_{21}^{AB}e^{ik_x}+t_{22}^{AB}e^{-ik_y}.
\end{aligned}
\label{ex1}
\end{equation}
The real-space TB model $H$ described by this Bloch Hamiltonian is depicted in Fig. \ref{fig:CLSexample}(b). 
The hopping directions that appear can be directly understood from the shape of the CLS, as illustrated in Fig. \ref{fig:CLSexample}(c): hoppings from site $A$ to site $B$ ($B$ to $A$, $A$ to $A$, $B$ to $B$) are determined by all vectors that connect orbitals $A$ to orbitals $B$ ($B$ to $A$, $B$ to $B$, $A$ to $A$) within the CLS (see Appendix \ref{AppA} for details). As a consequence, the model (\ref{ex1}) involves first- and second-neighbor hoppings.

The onsite energies
$V_A\equiv|w_{B_1}|^2+|w_{B_2}|^2$, $V_B\equiv|w_{A_1}|^2+|w_{A_2}|^2$, inter-sublattice hopping parameters \smash{$t_{ij}^{\alpha\beta}\equiv-w_{\alpha_i}w_{\beta_j}^*$} and intra-sublattice hopping parameters \smash{$t_{ij}^{\alpha\alpha}\equiv w_{\alpha_i}w_{\alpha_j}^*$} strongly depend on the CLS amplitudes. As a consequence, the band structure can be \emph{engineered} by the choice of the CLS amplitudes. In particular, the existence of a (singular) BTP can be ensured by choosing $\ket{f_\mathbf{k}}$ singular. For example, if the CLS amplitudes are such that $\ket{f_{\mathbf{k}_\text{X}}}=0$, a BTP at the X point of the FBZ appears, see Fig. \ref{fig:CLSexample}(d). Similarly, there would be a BTP at the $\Gamma$ point if \smash{$w_{\alpha_1}=-w_{\alpha_2}\in\mathbb{R}$}. In contrast, if $\ket{f_\mathbf{k}}$ is chosen non-singular, the bands are gapped away from each other, see Fig. \ref{fig:CLSexample}(e).

\begin{figure}
	\centering
	\includegraphics[width=\columnwidth]{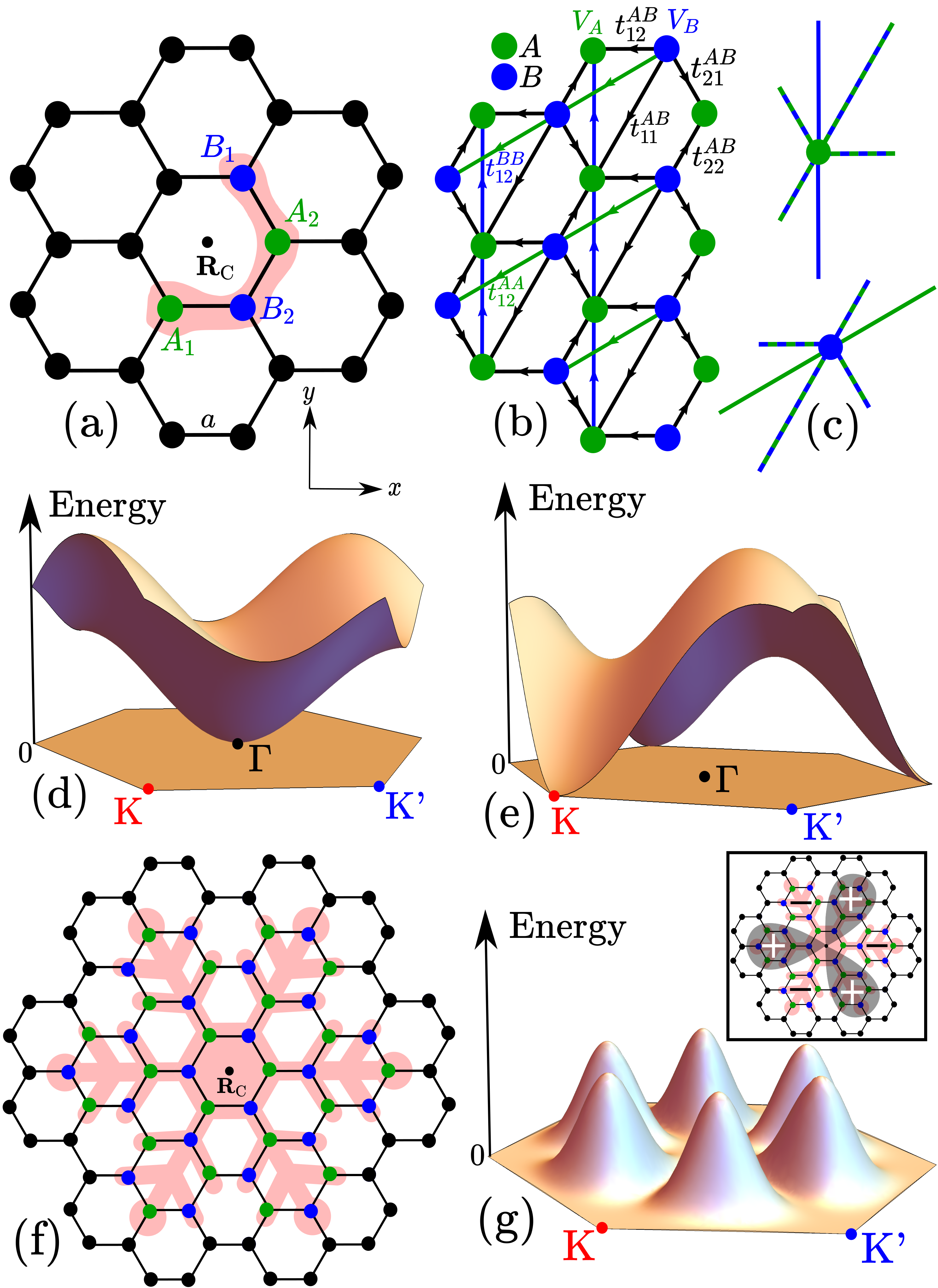}
	\caption{(a) CLS on the honeycomb lattice. (b) Flat-band TB model built from it. (c) The hopping directions are determined by the CLS shape. (d) Band structure for $w_{\alpha_1}=-w_{\alpha_2}=1$. (e) Band structure for $w_{\alpha_1}=1$, $w_{\alpha_2}=-e^{2\pi i/3}$. (f) Snowflake-shaped CLS and (g) the band structure of the corresponding flat-band TB model, with amplitudes as shown in the inset: \smash{$w_{\alpha_i}=1$} inside the dark-shaded branches, and \smash{$w_{\alpha_i}=-1$} outside.}
	\label{fig:twobandhex}
\end{figure}

 As a second example, consider the CLS on the honeycomb lattice shown in Fig. \ref{fig:twobandhex}(a). 
It is straightforward to obtain the BCLS
 \begin{equation}
\ket{f_\mathbf{k}}=\begin{pmatrix}
w_{A_1}e^{-ik_+}+w_{A_2}e^{ik_x}\\
w_{B_1}e^{ik_+}+w_{B_2}e^{ik_-}
\end{pmatrix},
 \end{equation}
 where again $k_\pm\equiv (k_x\pm \sqrt{3}k_y)/2$. Inserting into Eq. (\ref{2Ham}), one obtains a TB model of the form (\ref{expl2}), where now
 \begin{equation}
\begin{aligned}
H_{AA}&=V_A+t_{12}^{BB}e^{i\sqrt{3}k_y}+t_{21}^{BB}e^{-i\sqrt{3}k_y},\\
H_{BB}&=V_B+t_{12}^{AA}e^{-i(k_x+k_+)}+t_{21}^{AA}e^{i(k_x+k_+)},\\
H_{AB}&=t_{11}^{AB}e^{-2ik_+}+t_{12}^{AB}e^{-ik_x}\\
&+t_{21}^{AB}e^{ik_-}+t_{22}^{AB}e^{ik_+},
\end{aligned}
 \end{equation}
with onsite energies and hoppings as defined above. This TB model is shown in Fig. \ref{fig:twobandhex}(b). Due to the asymmetric shape of the input CLS, the hoppings are also distributed asymmetrically, as illustrated in Fig. \ref{fig:twobandhex}(c). By an appropriate choice of the CLS amplitudes, the band structure can be freely designed to exhibit a BTP at the $\Gamma$ point or at the K points, see Fig. \ref{fig:twobandhex}(d)\&(e), or to be gapped by taking $\ket{f_\mathbf{k}}$ non-singular.

 Proceeding in the same way, infinitely many different $N=2$ flat-band models and their band structures can be designed. For \emph{any imaginable CLS} (\ref{2CLS}), if one inserts the corresponding BCLS (\ref{2BCLS}) into Eq. (\ref{2Ham}), one obtains a flat-band TB model with hopping directions that are determined by the shape of the CLS (i.e. the set of all vectors \smash{$\boldsymbol{\delta}_{\alpha_i}$}), and hopping parameters as well as onsite energies that are determined by the CLS amplitudes $w_{\alpha_i}$.
 
As explained in more detail in Appendix \ref{AppA}, the Bloch Hamiltonian (\ref{2Ham}) is indeed always \emph{reasonable} [cf. condition (\ref{constr2})] due to its quadratic character.
In fact, it is not only a reasonable flat-band Hamiltonian, but the \emph{only possible form} for the Bloch Hamiltonian of any $N=2$ flat-band model. This was already anticipated by Rhim and Yang \cite{Rhim_2019} and can be easily understood from the spectral theorem, see Appendix \ref{app2der}.

 To illustrate that the procedure outlined above indeed works for arbitrarily complicated input CLSs, we may consider the snowflake-shaped CLS displayed in Fig. \ref{fig:twobandhex}(f). Of course, for such an extended input CLS, the real-space model that one obtains involves many further-neighbor hoppings and it is not useful to draw it explicitly. However, conveniently, the band structure (\ref{2bands}) can be analyzed and designed \emph{before} considering the real-space model. For example, taking $\smash{w_{\alpha_i}=1}$ for all 42 orbitals involved in the CLS, the dispersive band $\epsilon_1(\mathbf{k})$ strongly oscillates and exhibits a large peak at the $\Gamma$ point (not shown). If the amplitudes are flipped to $\smash{w_{\alpha_i}=-1}$ on three out of the six branches of the snowflake, see Fig. \ref{fig:twobandhex}(g), then $\epsilon_1(\mathbf{k})$ exhibits six peaks arrayed hexagonally in the FBZ. 
  
As we will show, an analogous procedure of flat-band model and band structure design is possible for $N>2$. Additionally, the tunability turns out to be considerably increased.

\subsection{Three-band models}
\label{3quad}

Consider now a CLS built on a lattice with three orbitals per unit cell:
\begin{equation}
	|\Psi_\text{CLS}^{\mathbf{R}_\text{C}}\rangle=\sum_{i\in\text{CLS}}(w_{A_i}\ket{A_i}+w_{B_i}\ket{B_i}+w_{C_i}\ket{C_i}).
	\label{3CLS}
\end{equation}
The corresponding BCLS has three components,
\begin{equation}
\ket{f_\mathbf{k}}=(f_A,f_B,f_C)^T.
\end{equation} 
We can now introduce \emph{three} states orthogonal to $\ket{f_\mathbf{k}}$, namely
\begin{equation} 
	\begin{aligned}
		\ket{f_\mathbf{k}^{AB}}&=(-f_B^*,f_A^*,0)^T,\\ \ket{f_\mathbf{k}^{AC}}&=(-f_C^*,0,f_A^*)^T,\\ \ket{f_\mathbf{k}^{BC}}&=(0,-f_C^*,f_B^*)^T,
	\end{aligned}
\label{orthstates}
\end{equation} 
with corresponding matrices
\begin{equation}
	\begin{aligned}
		F^{AB}_\mathbf{k}&\equiv\ket{f^{AB}_\mathbf{k}}\bra{f^{AB}_\mathbf{k}}=\begin{bmatrix}
			|f_B|^2 & -f_Af_B^* &0\\
			-f_A^*f_B & |f_A|^2 &0\\
			0&0&0
		\end{bmatrix},\\
		F^{AC}_\mathbf{k}&\equiv\ket{f^{AC}_\mathbf{k}}\bra{f^{AC}_\mathbf{k}}=\begin{bmatrix}
			|f_C|^2 & 0&-f_Af_C^* \\
			0&0&0\\
			-f_A^*f_C &0& |f_A|^2 
		\end{bmatrix},\\
		F^{BC}_\mathbf{k}&\equiv\ket{f^{BC}_\mathbf{k}}\bra{f^{BC}_\mathbf{k}}=\begin{bmatrix}
			0&0&0\\
			0&|f_C|^2 & -f_Bf_C^* \\
			0&-f_B^*f_C & |f_B|^2 \\
		\end{bmatrix}.
	\end{aligned}
\end{equation} 
A class of flat-band Bloch Hamiltonians is then obtained as a linear combination:
\begin{equation}
	H_\mathbf{k}=\lambda_\mathbf{k}^{AB} F^{AB}_\mathbf{k}+\lambda_\mathbf{k}^{AC} F^{AC}_\mathbf{k}+\lambda_\mathbf{k}^{BC} F^{BC}_\mathbf{k},
	\label{3Ham}
\end{equation}
where \smash{$\lambda_\mathbf{k}^{AB}$, $\lambda_\mathbf{k}^{AC}$, $\lambda_\mathbf{k}^{BC}$} are arbitrary real functions with the periodicity of the FBZ. It will suffice to choose them as constants, \smash{$\lambda_\mathbf{k}^{\alpha\beta}=\lambda_{\alpha\beta}$}.

The Hamiltonian (\ref{3Ham}) is analogous to Eq. (\ref{2Ham}) in that it automatically verifies the desired conditions (\ref{matcon}) \& (\ref{constr2}) for any conceivable input CLS, i.e. for any $\ket{f_\mathbf{k}}$. Therefore, infinitely many flat-band models on any $N=3$ lattice can be obtained from Eq. (\ref{3Ham}).

Moreover, there is an additional advantage compared to the $N=2$ case, namely there are now three \emph{relative} parameters $\lambda_{\alpha\beta}$, while the single parameter $\lambda_{AB}$ in the two-band Hamiltonian (\ref{2Ham}) only acts globally. As a consequence, the Hamiltonian (\ref{3Ham}) can be strongly tuned even if an input CLS \emph{completely fixed with regard to both shape and amplitudes} is used. This represents an additional degree of freedom absent in the two-band case.
  In this context, it is crucial to realize that the states (\ref{orthstates}) are \emph{not} eigenstates of the Hamiltonian (\ref{3Ham}). In fact, they form an overcomplete basis of the space orthogonal to $\ket{f_\mathbf{k}}$. This redundancy is at the origin of the tunability provided by the $\lambda_{\alpha\beta}$. In contrast to the flat-band eigenstate $\ket{f_\mathbf{k}}$, the dispersive-band eigenstates of the Hamiltonian (\ref{3Ham}) do depend on the parameters $\lambda_{\alpha\beta}$.

The dependency on the choice of the input CLS (i.e. of the functions $f_\alpha$) and the tunability by the $\lambda_{\alpha\beta}$ are clearly reflected in the Hamiltonian's band structure
\begin{equation}
	\begin{aligned}
		\epsilon_0(\mathbf{k})&=0,\\
		\epsilon_{1,2}(\mathbf{k})&=\frac{1}{2}\left(C_1\pm\sqrt{2C_2-C_1^2}\right),
	\end{aligned}
	\label{spec3}
\end{equation}
where $C_n\equiv \Tr(H^n_\mathbf{k})$ are given by
\begin{equation}
	\begin{aligned}
		C_1&\equiv(\lambda_{AB}+\lambda_{AC})|f_A|^2+(\lambda_{AB}+\lambda_{BC})|f_B|^2\\
		&+(\lambda_{AC}+\lambda_{BC})|f_C|^2,\\
		C_2&\equiv(\lambda_{AC}|f_A|^2+\lambda_{BC}|f_B|^2)^2\\
		&+(\lambda_{AB}|f_A|^2+\lambda_{BC}|f_C|^2)^2\\
		&+(\lambda_{AB}|f_B|^2+\lambda_{AC}|f_C|^2)^2\\
		&+2(\lambda_{AB}^2|f_A|^2|f_B|^2+\lambda_{AC}^2|f_A|^2|f_C|^2\\
		&\hspace{.5cm}+\lambda_{BC}^2|f_B|^2|f_C|^2).
	\end{aligned}
\end{equation}

\begin{figure}
	\centering
	\includegraphics[width=\columnwidth]{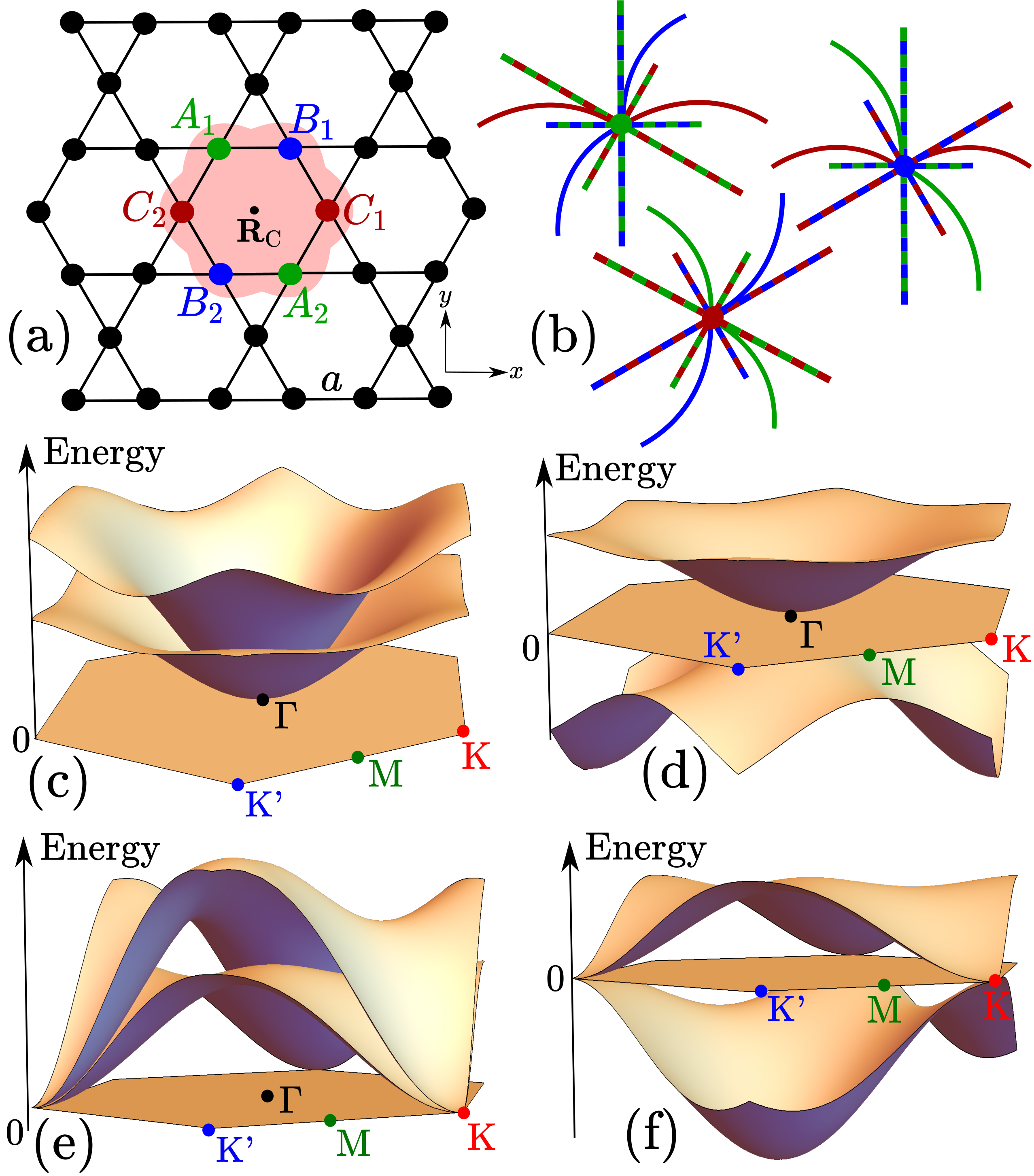}
	\caption{(a) CLS on the Kagome lattice. (b) Hopping directions occurring in the flat-band TB model built from it, as determined by the CLS shape. The corresponding band structure is shown for \smash{$w_{\alpha_1}=-w_{\alpha_2}=1$} in (c) with $\lambda_{AB}=\lambda_{AC}=\lambda_{BC}/2=1$ and in (d) with $\lambda_{AB}=\lambda_{AC}=-\lambda_{BC}/2=1$; for $w_{\alpha_1}=1$, $w_{A_2}=e^{\pi i/3}$, $w_{B_2}=e^{-\pi i/3}$, $w_{C_2}=-e^{-2\pi i/3}$ in (e) with $\lambda_{AB}=\lambda_{AC}=\lambda_{BC}/2=1$ and in (f) with $\lambda_{AB}=\lambda_{AC}=-\lambda_{BC}/2=1$.
	}
	\label{fig:kagome}
\end{figure}

As an example, consider again the six-site CLS on the Kagome lattice used previously, shown in Fig. \ref{fig:kagome}(a), with corresponding BCLS (\ref{kagomef}).
 Inserting into Eq. (\ref{3Ham}), one obtains a flat-band TB model
\begin{equation}
	H_\mathbf{k}=\begin{bmatrix}
		H_{AA} & H_{AB} &H_{AC}\\
		H_{AB}^* & H_{BB} &H_{BC}\\
		H_{AC}^* & H_{BC}^* & H_{CC}
	\end{bmatrix},
\label{TBmod}
\end{equation}
with diagonal matrix elements
\begin{equation}
	\begin{aligned}
		H_{AA}&=V_A+t_{12}^{A,BB}e^{2ik_+}+t_{21}^{A,BB}e^{-2ik_+}\\
		&+t_{12}^{A,CC}e^{2ik_x}+t_{21}^{A,CC}e^{-2ik_x},\\
		H_{BB}&=V_B+t_{12}^{B,AA}e^{-2ik_-}+t_{21}^{B,AA}e^{2ik_-}\\
		&+t_{12}^{B,CC}e^{2ik_x}+t_{21}^{B,CC}e^{-2ik_x},\\
		H_{CC}&=V_C+t_{12}^{C,AA}e^{-2ik_-}+t_{21}^{C,AA}e^{2ik_-}\\
		&+t_{12}^{C,BB}e^{2ik_+}+t_{21}^{C,BB}e^{-2ik_+},
	\end{aligned}
\end{equation}
and off-diagonal elements
\begin{equation}
\begin{aligned}
H_{AB}&=t_{11}^{AB}e^{-ik_x}+t_{12}^{AB}e^{i\sqrt{3}k_y}\\
&+t_{21}^{AB}e^{-i\sqrt{3}k_y}+t_{22}^{AB}e^{ik_x},\\
H_{AC}&=t_{11}^{AC}e^{-i(k_x+k_-)}+t_{12}^{AC}e^{ik_+}\\
&+t_{21}^{AC}e^{-ik_+}+t_{22}^{AC}e^{i(k_x+k_-)},\\
H_{BC}&=t_{11}^{BC}e^{-ik_-}+t_{12}^{BC}e^{i(k_x+k_+)}\\
&+t_{21}^{BC}e^{-i(k_x+k_+)}+t_{22}^{BC}e^{ik_-}.
\end{aligned}
\end{equation}
Again, the hopping directions are determined by the CLS shape, as illustrated in Fig. \ref{fig:kagome}(b), and the TB model thus involves at most third-neighbor hoppings.

The onsite energies \smash{$V_\alpha\equiv\sum_{\beta\neq\alpha}\lambda_{\alpha\beta}(|w_{\beta_1}|^2+|w_{\beta_2}|^2)$}, inter-sublattice hoppings \smash{$t_{ij}^{\alpha\beta}\equiv-\lambda_{\alpha\beta}w_{\alpha_i}w_{\beta_j}^*$} and intra-sublattice hoppings \smash{$t_{ij}^{\alpha,\beta\beta}\equiv\lambda_{\alpha\beta}w_{\beta_i}w_{\beta_j}^*$}  are strongly tunable by the six CLS amplitudes \smash{$w_{\alpha_i}$} and the three parameters $\lambda_{\alpha\beta}$.
This affords considerable control over the behavior of the dispersive bands, and in particular over the occurrence of BTPs. For example, the band structure can be chosen to exhibit a BTP at the $\Gamma$ point, and the position of the flat band with respect to the dispersive bands can be controlled by the $\lambda_{\alpha\beta}$, see Fig. \ref{fig:kagome}(c)\&(d). 
Similarly, the BTP can be placed at the K points, and again the position of the flat band can be controlled, see Fig. \ref{fig:kagome}(e)\&(f). Of course, the flat band can also be gapped out completely by an imbalance in the CLS amplitudes.

More generally, for three-band models of the form (\ref{3Ham}), a BTP of any degeneracy, and at various positions in the FBZ, can be designed without destroying the flat band. At $\mathbf{k}=\mathbf{k}_0$, there will be a twofold BTP if $2C_2(\mathbf{k}_0)=C_1(\mathbf{k}_0)[1+C_1(\mathbf{k}_0)]$, and a threefold BTP if $C_1(\mathbf{k}_0)=C_2(\mathbf{k}_0)=0$. 
Clearly, this can be achieved in several different ways, with or without fine-tuning of the $\lambda_{\alpha\beta}$. All possible scenarios are listed in Appendix \ref{3BTPs}. 

\subsection{$N$-band models} 
\label{Nquad}

The arguments made above can be readily extended to any $N$. Consider a generic CLS (\ref{CLSdef}) built on a lattice with $N$ orbitals per unit cell, with corresponding BCLS
\begin{equation}
 \ket{f_\mathbf{k}}=(f_A,f_B,f_C,...)^T. 
 \end{equation}
We may now introduce $\binom{N}{2}$ states 
\begin{equation}
	|f^{\alpha\beta}_\mathbf{k}\rangle\equiv(0,...,0,-f_\beta^*,0,...,0,f_\alpha,0,...,0)^T
	\label{Nstates}
\end{equation}
orthogonal to $\ket{f_\mathbf{k}}$, and the $N\times N$ matrices
\begin{equation}
	F^{\alpha\beta}_\mathbf{k}\equiv|f^{\alpha\beta}_\mathbf{k}\rangle\langle f^{\alpha\beta}_\mathbf{k}|
\end{equation}
constructed from them. A generic quadratic flat-band Bloch Hamiltonian is then obtained by forming a linear combination of these $\binom{N}{2}$ matrices:
\begin{equation}
	H_\mathbf{k}=\sum_{\alpha,\beta>\alpha}\lambda^{\alpha\beta}_\mathbf{k}F^{\alpha\beta}_\mathbf{k},
	\label{mainham}
\end{equation}
where $\alpha,\beta\in\{A,B,C,...\}$ and \smash{$\lambda^{\alpha\beta}_\mathbf{k}$} is an arbitrary real function with the periodicity of the FBZ. We will assume \smash{$\lambda^{\alpha\beta}_\mathbf{k}=\lambda_{\alpha\beta}$}, such that the off-diagonal and diagonal matrix elements of the Hamiltonian (\ref{mainham}) explicitly read
\begin{equation}
	\begin{aligned}
		H_{\mathbf{k},\alpha\beta}&=-\lambda_{\alpha\beta}f_\alpha f_\beta^*,\\
		H_{\mathbf{k},\alpha\alpha}&=\sum_{\beta\neq\alpha}\lambda_{\alpha\beta}|f_\beta|^2.
\end{aligned}
\label{matelsquad}
\end{equation}

The Hamiltonian (\ref{mainham}), which generalizes Eqs. (\ref{2Ham}) \& (\ref{3Ham}) to any $N$, is the first main result of this paper. One can insert into it \emph{any arbitrary CLS} built on any lattice, for any spatial dimension and any number of sites per unit cell. Since conditions (\ref{matcon}) \& (\ref{constr2}) are automatically fulfilled, each such input CLS gives rise to a reasonable real-space tight-binding model with a flat band. One thus obtains arbitrarily many flat-band models on any lattice. 

Of course, while Eq. (\ref{mainham}) provides infinitely many flat-band models, it does not capture all possible flat-band models except in the $N=2$ limit. In particular, flat-band models with a linear function $H_\mathbf{k}(\ket{f_\mathbf{k}})$ exist for $N\geq3$, see Section \ref{linmods}.

 The flat-band models obtained from Eq. (\ref{mainham}) are tunable by two independent knobs: the functions $f_\alpha$ and the $N(N-1)/2$ parameters $\lambda_{\alpha\beta}$. The latter tunability stems from the fact that the states (\ref{Nstates}) are not eigenstates of the Hamiltonian (\ref{mainham}) and form an overcomplete basis of the space orthogonal to $\ket{f_\mathbf{k}}$.
 
  Although, for $N>3$, there is no elegant closed-form solution for the energy bands, the existence of a flat band is always guaranteed, and BTPs with any degree of degeneracy ($d=1,...,N$) can be designed.
The two most interesting classes of BTPs (not requiring fine-tuning of the $\lambda_{\alpha\beta}$) are the following. (1) If $\ket{f_{\mathbf{k}_0}}=0$ for some $\mathbf{k}_0$, there will be an $N$-fold (singular) BTP at $\mathbf{k}_0$. We have already seen examples for this in the two-band and three-band case, cf. Figs. \ref{fig:CLSexample}--\ref{fig:kagome}.
(2) If, at $\mathbf{k}_0$, all except one component of $\ket{f_{\mathbf{k}_0}}$ vanish, say $f_{\alpha_0}(\mathbf{k}_0)\neq0$ and $f_{\alpha\neq\alpha_0}(\mathbf{k}_0)=0$, then the Hamiltonian (\ref{mainham}) becomes locally diagonal with eigenvalues $\{0,\lambda_{\alpha_0,\alpha\neq\alpha_0}|f_{\alpha_0}(\mathbf{k}_0)|^2\}$. In other words, a $d$-fold BTP at $\mathbf{k}_0$, with $1\leq d\leq N-1$, can be created as desired by setting $d-1$ parameters $\lambda_{\alpha_0,\alpha\neq\alpha_0}$ to zero.

\begin{figure}
	\centering
	\includegraphics[width=\columnwidth]{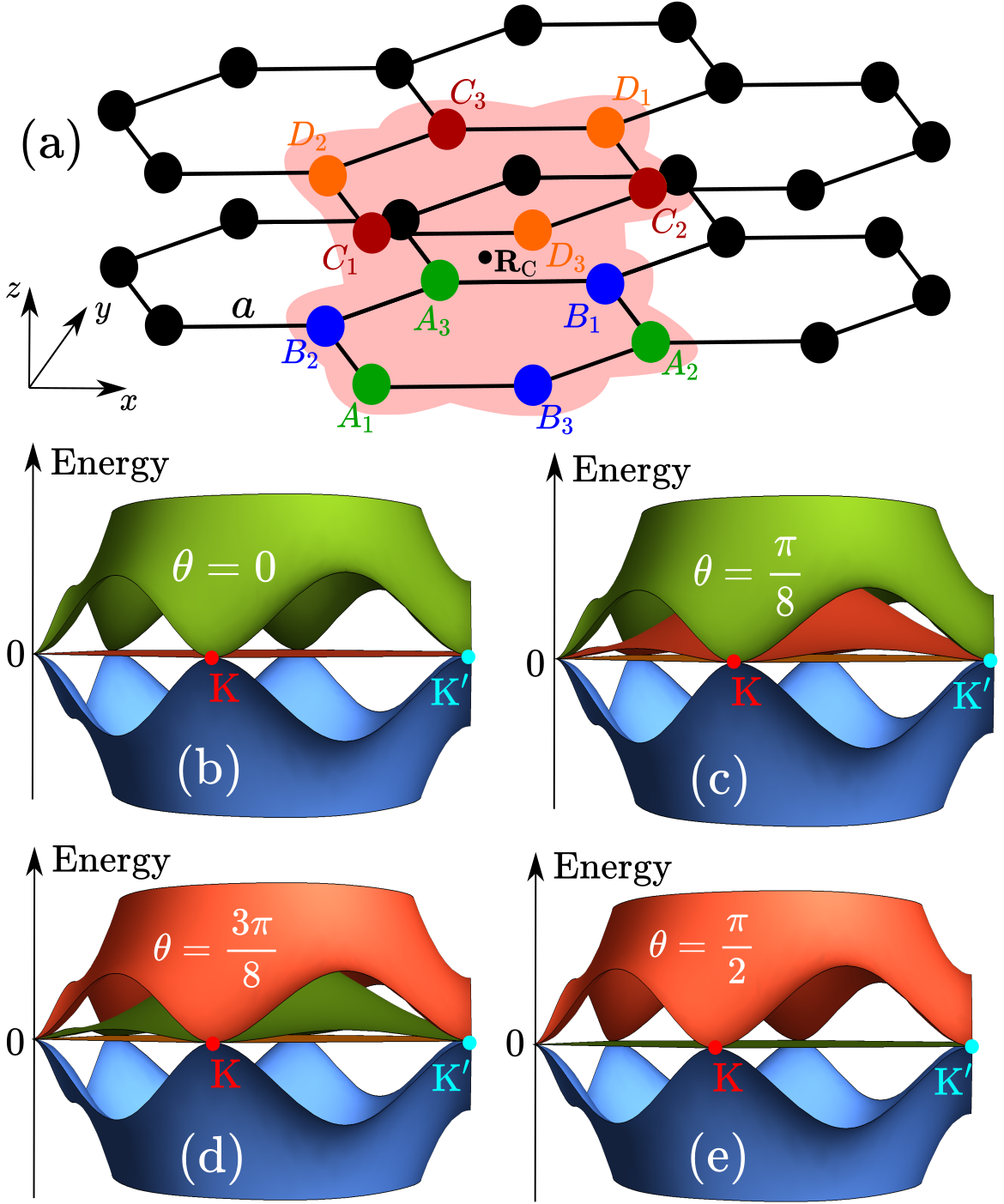}
	\caption{(a) CLS on a bilayer honeycomb lattice. (b)--(e) Band structure (\ref{bands4}) for the TB model built from it, for different values of $\theta$ and with $\lambda_{AB}=1$.}
	\label{fig:bilayer}
\end{figure}

As an example for the $N>3$ case, consider a CLS on an $N=4$ bilayer honeycomb lattice, shown in Fig. \ref{fig:bilayer}(a).
 For simplicity, we take all CLS amplitudes to be identical, \smash{$w_{\alpha_i}=1$}.
 The corresponding BCLS is given by 
 \begin{equation}
 \begin{aligned}
 \ket{f_\mathbf{k}}&=(g_\mathbf{k},g^*_\mathbf{k},g_\mathbf{k},g^*_\mathbf{k})^T,\\
 g_\mathbf{k}&\equiv2e^{-\frac{i}{2}k_x}\cos(\sqrt{3}k_y/2)+e^{ik_x}.
 \end{aligned}
 \end{equation}
 Even though the CLS is completely fixed (shape and amplitudes), the flat-band Bloch Hamiltonian (\ref{mainham}) built from it is still tunable by six parameters $\lambda_{AB}, \lambda_{AC},\lambda_{AD},\lambda_{BC},\lambda_{BD},\lambda_{CD}$. 
 To reduce the number of parameters, it is reasonable to set $\lambda_{CD}=\lambda_{AB}$, $\lambda_{BC}=\lambda_{AD}$ and $\lambda_{BD}=\lambda_{AC}$ for reasons of symmetry [cf. Fig. \ref{fig:bilayer}(a)]. We may then further set $\lambda_{AC}=-c_\theta^2\lambda_{AB}$ and $\lambda_{AD}=-s_\theta^2\lambda_{AB}$ to ensure a tight-binding model with zero onsite energies and no intra-sublattice hopping: 
\begin{equation}
H_\mathbf{k}=\lambda_{AB}\begin{bmatrix}
0&-g^2&c_\theta^2|g|^2&s_\theta^2g^2\\
-(g^*)^2 & 0 & s_\theta^2(g^*)^2 & c_\theta^2|g|^2\\
c_\theta^2|g|^2&s_\theta^2g^2&0& -g^2\\
s_\theta^2(g^*)^2  & c_\theta^2|g|^2 & -(g^*)^2 & 0
\end{bmatrix}.
\label{bilham}
\end{equation}
Here, the shorthand notation $s_\theta\equiv\sin\theta$ and $c_\theta\equiv\cos\theta$ is used.
Physically, the parameter $\theta$ quantifies the ratio between vertical ($AC$ and $BD$) and diagonal ($AD$ and $BC$) inter-layer hoppings.  
The band structure of the model (\ref{bilham}) reads
\begin{equation}
	\begin{aligned}
\epsilon_0(\mathbf{k})&=0, && \epsilon_1(\mathbf{k})&=-2|g|^2,\\
\epsilon_2(\mathbf{k})&=2s_\theta^2|g|^2, &&
\epsilon_3(\mathbf{k})&=2c_\theta^2|g|^2,
\end{aligned}
\label{bands4}
\end{equation} 
in units of $\lambda_{AB}$. The upper two (lower two) bands are tunable by $\theta$ if $\lambda_{AB}$ is taken to be positive (negative), as shown in Fig. \ref{fig:bilayer}(b)--(e). Note that the system has a chiral symmetry for $\theta=n\pi/2$, $n\in\mathbb{Z}$. In this case there is a doubly degenerate flat band, of which one is essential ($\epsilon_0$) while the other one is accidental ($\epsilon_2$ or $\epsilon_3$).

Finally, to illustrate that our flat-band model construction scheme indeed works for any dimension, an $N=4$ flat-band model built from a CLS on the (3D) cubic lattice is discussed in Appendix \ref{Appcube}.

\section{Flat-band models with linear band touching}
\label{linmods}

In the previous Section we have seen that Bloch Hamiltonians which are purely quadratic functions of the  components $f_\alpha$ of the flat-band eigenstate give rise to quadratic BTPs. It it then natural to study Bloch Hamiltonians that are purely linear functions of the  components $f_\alpha$. As might be expected, this will give rise to models with linear BTPs. 

The simplest scenario for a linear flat-band model is encountered in three-band ($N=3$) systems (Section \ref{lin3exs}). This case will be studied in detail before generalizing to higher $N$ (Section \ref{linconstr}).

\subsection{Three-band models}
\label{lin3exs}

Consider again some CLS built on a lattice with three orbitals per unit cell:
\begin{equation}
|\Psi_\text{CLS}^{\mathbf{R}_\text{C}}\rangle=\sum_{i\in\text{CLS}}(w_{A_i}\ket{A_i}+w_{B_i}\ket{B_i}+w_{C_i}\ket{C_i}),
	\label{3CLSb}
\end{equation}
with corresponding BCLS
\begin{equation}
	\ket{f_\mathbf{k}}=(f_A,f_B,f_C)^T.
	\label{3BCLSb}
\end{equation} 
Our goal is now to construct $N=3$ flat-band Bloch Hamiltonians $H_\mathbf{k}$ that are \emph{linear functions} of $\ket{f_\mathbf{k}}$. To achieve this, one cannot use a completely arbitrary input CLS, in contrast to the quadratic scenario. Instead, the CLS needs to have special properties. 

We can identify two classes of CLSs that allow to build a linear $H_\mathbf{k}$. More specifically, CLSs belonging to the first class may be called \emph{chiral CLSs}, since the flat-band Bloch Hamiltonians built from them are characterized by an effective (global) chiral symmetry \cite{Ryu_2010}
\begin{equation}
\mathcal{S}^\dagger H_\mathbf{k}\mathcal{S}=-H_\mathbf{k},
\label{chiral}
\end{equation}
with a unitary matrix $\mathcal{S}$ that fulfills $\mathcal{S}^2=\mathbf{1}_3$.  As it turns out, in the class of chiral CLSs we can further distinguish between a \emph{type-I chiral CLS (C$_I$-CLS)}, in which case $f_Af_Bf_C=0$, and a \emph{type-II chiral CLS (C$_{II}$-CLS)}, in which case $f_Af_Bf_C\neq0$. Similarly, CLSs belonging to the second class may be called \emph{CP-CLSs}, since they give rise to flat-band Bloch Hamiltonians that exhibit an effective (global) CP symmetry
\begin{equation}
\mathcal{C}^\dagger H_\mathbf{k}\mathcal{C}=-H_\mathbf{k}^*,
\label{chargecon}
\end{equation}
i.e. a combined symmetry of charge conjugation (C), defined by $\mathcal{V}^\dagger H_\mathbf{k}\mathcal{V}=-H_{-\mathbf{k}}^*$, and inversion [or parity (P)], defined by $\mathcal{P}^\dagger H_\mathbf{k}\mathcal{P}=H_{-\mathbf{k}}$ \cite{Ryu_2010}. Here, $\mathcal{C}=\mathcal{PV}$ is a unitary matrix that fulfills $\mathcal{C}^2=\mathbf{1}_3$. In the class of CP-CLSs one can further distinguish between a \emph{type-I CP-CLS (CP$_I$-CLS)}, in which case $f_Af_Bf_C=-(f_Af_Bf_C)^*$, and a \emph{type-II CP-CLS (CP$_{II}$-CLS)}, in which case $f_Af_Bf_C=(f_Af_Bf_C)^*$.

These four distinct generic classes of CLSs are presented in detail in the following. Note however that it is not clear whether they constitute an exhaustive set for the construction of $N=3$ linear flat-band models. It may be rewarding to clarify this question in the future.

\subsubsection{Flat-band models from chiral CLSs}

\subsubsection*{Type-I chiral CLSs}

The first possibility for building an $N=3$ linear flat-band Hamiltonian arises if the input CLS (\ref{3CLSb}) occupies two sublattices while vanishing on the third. Let $\tau\in\{A,B,C\}$ denote the unoccupied sublattice, then $f_\tau=0$ and $f_{\alpha\neq\tau}\neq0$. In this case we may call the CLS \emph{type-I chiral (C$_I$)}, and we have $f_Af_Bf_C=0$. 
Any C$_I$-CLS gives rise to a \emph{C$_I$-BCLS}
 \begin{equation}
 	\begin{aligned}
 		\ket{f_\mathbf{k}}&=|m_{\tau,\mathbf{k}}^{ABC}\rangle,\\
 		|m_{A,\mathbf{k}}^{ABC}\rangle&\equiv(0,f_B,f_C)^T,\\
 		|m_{B,\mathbf{k}}^{ABC}\rangle&\equiv(f_A,0,f_C)^T,\\
 	 |m_{C,\mathbf{k}}^{ABC}\rangle&\equiv(f_A,f_B,0)^T.
 	\end{aligned}
 \label{chiralBCLS}
 \end{equation}
It is easy to see that the matrix \smash{$M_{\tau,\mathbf{k}}^{ABC}$}, where
  \begin{equation}
 	\begin{aligned}
 		M_{A,\mathbf{k}}^{ABC}&\equiv\begin{bmatrix}
 			0&-f_C&f_B\\
 			-f_C^*&0&0\\
 			f_B^* & 0 & 0
 		\end{bmatrix},\\
 		M_{B,\mathbf{k}}^{ABC}&\equiv\begin{bmatrix}
 			0 & -f_C^* & 0\\-f_C & 0 & f_A\\ 0 & f_A^* & 0
 		\end{bmatrix},\\	M_{C,\mathbf{k}}^{ABC}&\equiv\begin{bmatrix}
 			0&0&-f_B^*\\
 			0&0&f_A^*\\
 			-f_B & f_A & 0
 		\end{bmatrix},
 	\end{aligned}
 	\label{liebham}
 \end{equation}
vanishes on \smash{$|m_{\tau,\mathbf{k}}^{ABC}\rangle$}.
Consequently, any C$_I$-CLS can be used to build a linear flat-band Bloch Hamiltonian
\begin{equation}
	H_\mathbf{k}=\lambda_\mathbf{k}^{ABC}M_{\tau,\mathbf{k}}^{ABC},
	\label{HABC}
\end{equation} 
where \smash{$\lambda_\mathbf{k}^{ABC}$} is any function with the periodicity of the FBZ. This Hamiltonian has a band structure
\begin{equation}
	\begin{aligned}
		\epsilon_0(\mathbf{k})&=0,\\
		\epsilon_{1,2}(\mathbf{k})&=\pm\lambda_\mathbf{k}^{ABC}\sqrt{\langle m_{\tau,\mathbf{k}}^{ABC}|m_{\tau,\mathbf{k}}^{ABC}\rangle}.
	\end{aligned}
\label{bands3a}
\end{equation} 
The bands are particle-hole symmetric due to a chiral symmetry (\ref{chiral}) with $\mathcal{S}=\text{diag}(\sigma_A,\sigma_B,\sigma_C)$, where $\sigma_{\alpha\neq\tau}=1$ and $\sigma_{\alpha=\tau}=-1$. Note that \smash{$\lambda_\mathbf{k}^{ABC}=1$} in the following.

\begin{figure}
	\centering
	\includegraphics[width=\columnwidth]{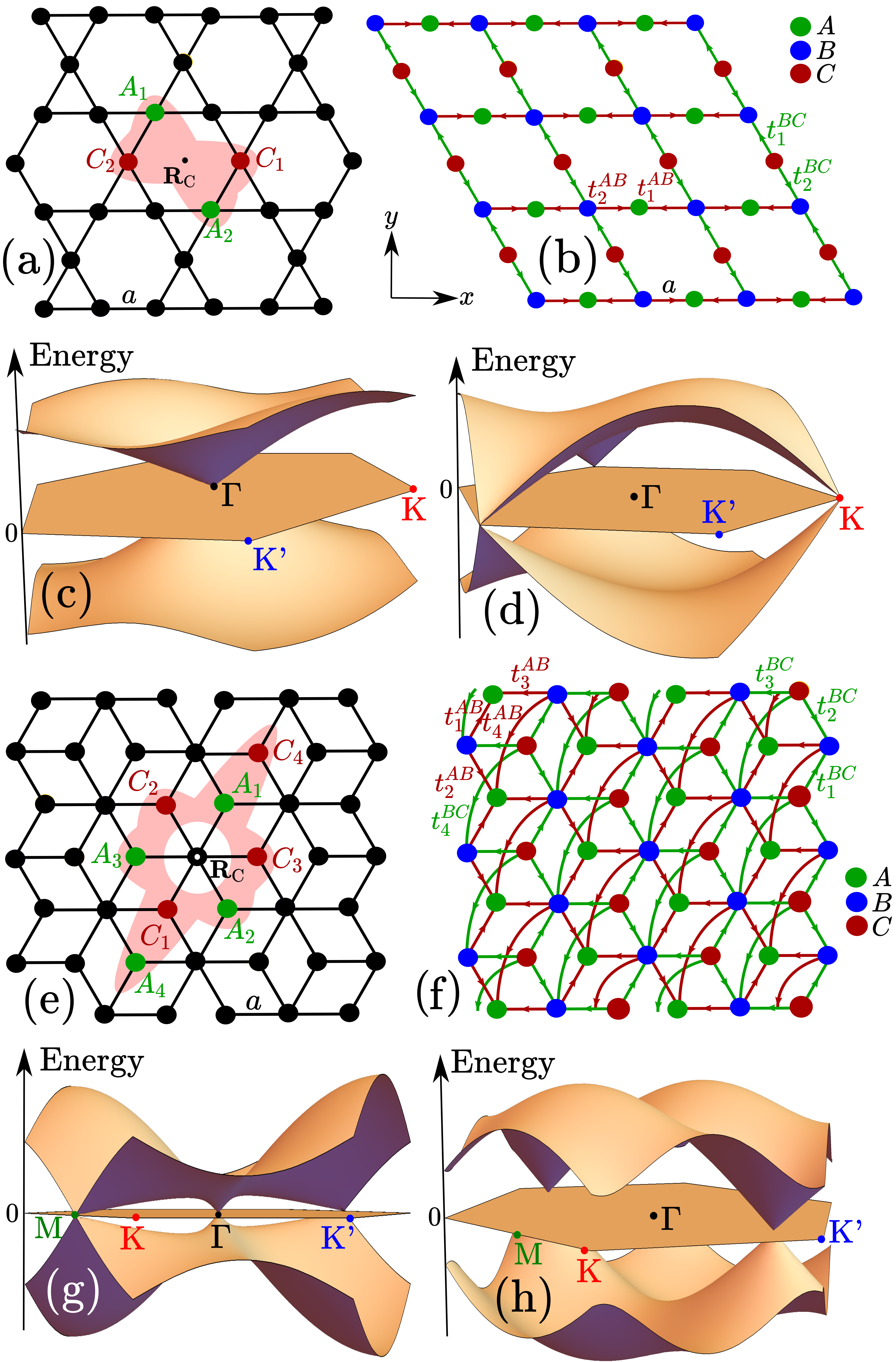}
	\caption{(a) A type-I chiral CLS on the Kagome lattice. (b) Flat-band TB model built from it. Hopping directions are determined by the CLS shape as indicated by the colors. The corresponding band structure is shown in (c) for $w_{\alpha_1}=-w_{\alpha_2}=1$ and in (d) for $w_{\alpha_1}=1$, $w_{\alpha_2}=e^{\pi i/3}$. (e) A more complicated C$_I$-CLS on the dice lattice. (f) TB model built from it. The corresponding band structure is shown in (g) for $w_{\alpha_1}=-w_{\alpha_4}=w_{A_2}=w_{C_2}/2=-w_{A_3}=-w_{C_3}/2=1$ and in (h) for $w_{\alpha_{1,2,3}}=-w_{\alpha_4}=1$.}
	\label{fig:linearexs}
\end{figure}

Type-I chiral CLSs can be easily found on any $N=3$ lattice, since the functions $f_{\alpha\neq\tau}$ are completely unconstrained. As an example, consider the CLS on the Kagome lattice shown in Fig. \ref{fig:linearexs}(a).
 Evidently, we have $\tau=B$, and the associated BCLS (\ref{chiralBCLS}) is given by
\begin{equation}
\begin{aligned}
			f_A&=w_{A_1}e^{-i k_-}+w_{A_2}e^{i k_-},\\
f_C&=w_{C_1}e^{ik_x}+w_{C_2}e^{-ik_x}.\\
\end{aligned}
		\label{chiralf0}
\end{equation}
 Inserting into Eq. (\ref{HABC}), we obtain a flat-band tight-binding model
\begin{equation}
H_\mathbf{k}=\begin{bmatrix}
0 & H_{AB} & 0\\
H_{AB}^* & 0 & H_{BC}\\
0 & H_{BC}^* & 0
\end{bmatrix},
\label{f3ham}
\end{equation}
where
\begin{equation}
	\begin{aligned}
		H_{AB}&=t^{AB}_1e^{-ik_x}+t^{AB}_2e^{ik_x},\\
		H_{BC}&=t^{BC}_1e^{-ik_-}+t^{BC}_2e^{ik_-}.
	\end{aligned}
\end{equation} 
This model has the topology of a Lieb lattice and is depicted in Fig. \ref{fig:linearexs}(b). Again, the hopping directions are determined by the shape of the CLS, i.e. by the vectors \smash{$\boldsymbol{\delta}_{\alpha_i}$}, although in a different fashion than for the quadratic models (see Appendix \ref{AppA}). Similarly, the hopping parameters $t^{AB}_i\equiv-w_{C_i}^*$ and $t^{BC}_i\equiv w_{A_i}$ again depend on the CLS amplitudes, which allows to design the band structure. For instance, a (singular) BTP may be created at the $\Gamma$ point [Fig. \ref{fig:linearexs}(c)] or at the K points [Fig. \ref{fig:linearexs}(d)], while the flat band can be gapped out by an imbalance in the CLS amplitudes.

As a more complicated example, consider the C$_I$-CLS on the dice lattice shown in Fig. \ref{fig:linearexs}(e). The corresponding BCLS (\ref{chiralBCLS}) is given by
\begin{equation}
\begin{aligned}
f_A&=w_{A_1}e^{ik_+}+w_{A_2}e^{ik_-}\\
&+w_{A_3}e^{-ik_x}+w_{A_4}e^{-2ik_+},\\
f_C&=w_{C_1}e^{-ik_+}+w_{C_2}e^{-ik_-}\\
&+w_{C_3}e^{ik_x}+w_{C_4}e^{2ik_+},
\end{aligned}
\end{equation}
and inserting into Eq. (\ref{HABC}) leads to a flat-band TB model of the form (\ref{f3ham}), where now
\begin{equation}
\begin{aligned}
H_{AB}&=t_1^{AB}e^{ik_+}+t_2^{AB}e^{ik_-}\\
&+t_3^{AB}e^{-ik_x}+t_4^{AB}e^{-2ik_+},\\
H_{BC}&=t_1^{BC}e^{ik_+}+t_2^{BC}e^{ik_-}\\
&+t_3^{BC}e^{-ik_x}+t_4^{BC}e^{-2ik_+},
\end{aligned}
\end{equation}
with hopping parameters as defined above. This model is depicted in Fig. \ref{fig:linearexs}(f) and allows for different kinds of BTPs depending on the CLS amplitudes. For example, BTPs may appear at the $\Gamma$ and M points  [Fig. \ref{fig:linearexs}(g)], away from high-symmetry points [Fig. \ref{fig:linearexs}(h)], or the flat band may be gapped out by a suitable asymmetry.

Finally, note that a similar construction works in any spatial dimension, since C$_I$-CLSs can be easily built.

\subsubsection*{Type-II chiral CLSs}

The second possibility for building an $N=3$ linear flat-band Hamiltonian arises if the input CLS (\ref{3CLSb}) occupies all sublattices, with the following constraints: the orbitals of two types need to be located at the same positions and occupied with opposite CLS amplitudes. Further, the orbitals of the third sublattice (indicated by $\tau$) need to be placed pairwise at equal distance from the localization center, and the CLS amplitudes within each pair have to be correlated in an antisymmetrical fashion. This reflects the general property that the position of the localization center is not arbitrary for linear FB models (see Appendix \ref{AppA}, in particular Fig. \ref{fig:quadlindiff}). We may call a CLS of this kind \emph{type-II chiral (C$_{II}$)}. The constraints defining a C$_{II}$-CLS can be formulated more precisely in reciprocal space: any C$_{II}$-CLS corresponds to a \emph{C$_{II}$-BCLS}
\begin{equation}
\begin{aligned}
f_\mathbf{k}&=|n_{\tau,\mathbf{k}}^{ABC}\rangle,\\
|n_{A,\mathbf{k}}^{ABC}\rangle&=(f_A,-f_C,f_C)^T,\\
|n_{B,\mathbf{k}}^{ABC}\rangle&=(f_A,f_B,-f_A)^T,\\
|n_{C,\mathbf{k}}^{ABC}\rangle&=(-f_B,f_B,f_C)^T,
\label{BCLSCII}
\end{aligned}
\end{equation}
where the component $f_\tau$ is imaginary.
It is easy to see that the matrix \smash{$N_{\tau,\mathbf{k}}^{ABC}$}, where
\begin{equation}
	\begin{aligned}
		N_{A,\mathbf{k}}^{ABC}&\equiv\begin{bmatrix}
			0&f_C^*&f_C^*\\
			f_C&0&-f_A\\
			f_C & f_A & 0
		\end{bmatrix},\\
		N_{B,\mathbf{k}}^{ABC}&\equiv\begin{bmatrix}
			0 & f_A & f_B\\
			f_A^* & 0 & f_A^*\\ 
			-f_B & f_A & 0
		\end{bmatrix},\\	N_{C,\mathbf{k}}^{ABC}&\equiv\begin{bmatrix}
			0&-f_C&f_B\\
			f_C&0&f_B\\
			f_B^* & f_B^* & 0
		\end{bmatrix},
	\end{aligned}
\end{equation}
vanishes on \smash{$|n_{\tau,\mathbf{k}}^{ABC}\rangle$}.
Consequently, any C$_{II}$-CLS can be used to build a linear flat-band Bloch Hamiltonian
\begin{equation}
	H_\mathbf{k}=\lambda_\mathbf{k}^{ABC}N_{\tau,\mathbf{k}}^{ABC},
	\label{HABC2}
\end{equation} 
where \smash{$\lambda_\mathbf{k}^{ABC}$} is any function with the periodicity of the FBZ. This Hamiltonian has a band structure
\begin{equation}
	\begin{aligned}
		\epsilon_0(\mathbf{k})&=0,\\
		\epsilon_{1,2}(\mathbf{k})&=\pm\lambda_\mathbf{k}^{ABC}\sqrt{\langle n_{\tau,\mathbf{k}}^{ABC}|n_{\tau,\mathbf{k}}^{ABC}\rangle},
	\end{aligned}
	\label{bandsCII}
\end{equation} 
with the particle-hole symmetric character of the bands being protected by a chiral symmetry (\ref{chiral}) with diagonal matrix elements $\mathcal{S}_{\alpha\alpha}=-\delta_{\alpha\tau}$ and off-diagonal elements $\mathcal{S}_{\alpha\beta}=1-\delta_{\alpha\tau}-\delta_{\beta\tau}$. Again \smash{$\lambda_\mathbf{k}^{ABC}=1$} in the following.

\begin{figure}
	\centering
	\includegraphics[width=\columnwidth]{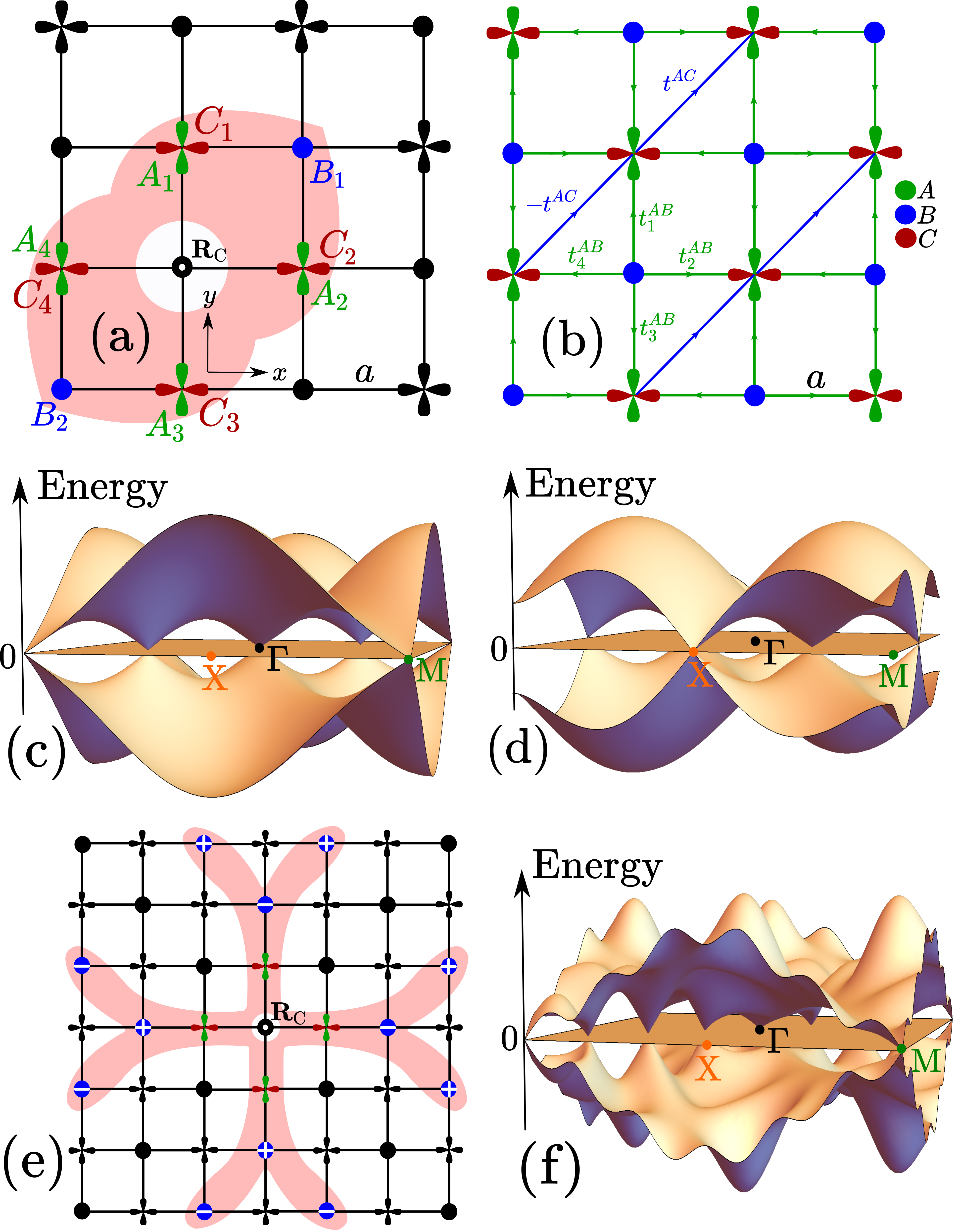}
	\caption{(a) A CLS on a multiorbital square lattice that is type-II chiral if the amplitudes are properly correlated. (b) Flat-band tight-binding model built from it. The corresponding band structure with $w_{B_1}=1$ is shown in (c) for $w_{A_1}=-w_{A_2}=w_{A_3}/2=-w_{A_4}/2=1$ and in (d) for $w_{A_1}=w_{A_2}=-w_{A_3}/2=-w_{A_4}/2=1$. (e) A more complicated C$_{II}$-CLS with $w_{A_1}=-w_{A_2}=w_{A_3}/2=-w_{A_4}/2=1$ and $w_{B_i}=\pm1$ as shown. (f) Band structure for the TB model built from it.}
	\label{fig:CII}
\end{figure}

Type-II chiral CLSs cannot be found on all $N=3$ lattices, due to the constraints on the CLS mentioned above. For example, a C$_{II}$-CLS cannot exist on the Kagome lattice, but on a multiorbital square lattice. As an example, consider the CLS shown in Fig.  \ref{fig:CII}(a), where we fix the amplitudes $w_{B_i}$ to be real for concreteness.
 In order for the CLS to be of C$_{II}$-type, we have to take $w_{A_i}=-w_{C_i}$ and $w_{B_1}=-w_{B_2}$, such that $\tau=B$. The corresponding BCLS (\ref{BCLSCII}) is then given by 
\begin{equation}
\begin{aligned}
f_A&=w_{A_1}e^{ik_y}+w_{A_2}e^{ik_x}\\
&+w_{A_3}e^{-ik_y}+w_{A_4}e^{-ik_x},\\
f_B&=2i w_{B_1}\text{sin}(k_x+k_y).
\end{aligned}
\end{equation}
 Inserting into Eq. (\ref{HABC2}), we obtain a flat-band tight-binding model
\begin{equation}
	H_\mathbf{k}=\begin{bmatrix}
		0 & H_{AB} & H_{AC}\\
		H_{AB}^* & 0 & H_{BC}\\
		H_{AC}^* & H_{BC}^* & 0
	\end{bmatrix},
	\label{CIImod}
\end{equation}
where
\begin{equation}
	\begin{aligned}
		H_{AB}&=H_{BC}^*=t^{AB}_1e^{ik_y}+t^{AB}_2e^{ik_x}\\
		&\hspace{1.2cm}+t^{AB}_3e^{-ik_y}+t^{AB}_4e^{-ik_x},\\
		H_{AC}&=t^{AC}e^{i(k_x+k_y)}-t^{AC}e^{-i(k_x+k_y)},
	\end{aligned}
\end{equation}
with $t^{AB}_i\equiv w_{A_i}$ and $t^{AC}\equiv w_{B_1}$, as shown in Fig.  \ref{fig:CII}(b). By choice of the CLS amplitudes a threefold BTP may be created for example at the $\Gamma$ and M points [Fig. \ref{fig:CII}(c)] or at the X point [Fig. \ref{fig:CII}(d)], as well as away from high-symmetry points.

A more complicated example for a C$_{II}$-CLS is shown in Fig. \ref{fig:CII}(e). Inserting into Eq. (\ref{HABC2}) leads to a flat-band TB model with linear BTPs at the $\Gamma$ and M points, as well as additonal BTPs along the (1,1)-direction, as shown in Fig. \ref{fig:CII}(f).

Note finally that C$_{II}$-CLSs can also be constructed in 3D, for example on a multiorbital cubic lattice.

\subsubsection{Flat-band models from CP-CLSs}

\subsubsection*{Type-I CP-CLSs}

The third possibility for building an $N=3$ linear flat-band Hamiltonian arises if the input CLS (\ref{3CLSb}) occupies all sublattices, with the following constraints: the orbitals of each type need to be arranged pairwise at equal distance from the localization center, and the signs of the CLS amplitudes $w_{\alpha_i}$ within each pair have to be correlated in a certain (anti)symmetrical fashion with respect to the localization center. Again, this reflects the important role of the localization center for linear FB models (see Appendix \ref{AppA}). We may call such a CLS \emph{type-I CP (CP$_I$)}.
More precisely, any CP$_I$-CLS corresponds to a \emph{CP$_I$-BCLS}
\begin{equation}
	\begin{aligned}
		\ket{f_\mathbf{k}}&=|o_\mathbf{k}^{ABC}\rangle\equiv(f_A,f_B,f_C)^T,\\
		f_\alpha^*&=\kappa_\alpha f_\alpha,\\ 
		\kappa_A\kappa_B\kappa_C&=-1,
	\end{aligned}
	\label{conjBCLS}
\end{equation}
i.e. each component $f_\alpha$ is either real ($\kappa_\alpha=1$) or imaginary ($\kappa_\alpha=-1$), and the three signs $\kappa_\alpha$ are correlated such that $f_Af_Bf_C=-(f_Af_Bf_C)^*$.
 It is easy to see that the matrix 
\begin{equation}
O_\mathbf{k}^{ABC}\equiv\begin{bmatrix}
0&-f_C&f_B\\
-\kappa_C f_C&0&\kappa_C f_A\\
\kappa_B f_B & -\kappa_B f_A & 0
\end{bmatrix}
\end{equation}
vanishes on \smash{$|o_\mathbf{k}^{ABC}\rangle$}. Thus, any CP$_I$-CLS can be used to build a linear flat-band Bloch Hamiltonian
\begin{equation}
	H_\mathbf{k}=\lambda_\mathbf{k}^{ABC}O_\mathbf{k}^{ABC}.
\label{linim}
\end{equation} 
This Hamiltonian has a band structure
\begin{equation}
	\begin{aligned}
		\epsilon_0(\mathbf{k})&=0,\\
		\epsilon_{1,2}(\mathbf{k})&=\pm\lambda_\mathbf{k}^{ABC}\sqrt{\langle o_\mathbf{k}^{ABC}|o_\mathbf{k}^{ABC}\rangle},
	\end{aligned}
	\label{bands3b}
\end{equation} 
which is particle-hole symmetric due to a CP symmetry (\ref{chargecon}) with $\mathcal{C}=\text{diag}(\kappa_A,\kappa_B,\kappa_C)$. Note that again \smash{$\lambda_\mathbf{k}^{ABC}=1$} hereafter.

\begin{figure}
	\centering
	\includegraphics[width=\columnwidth]{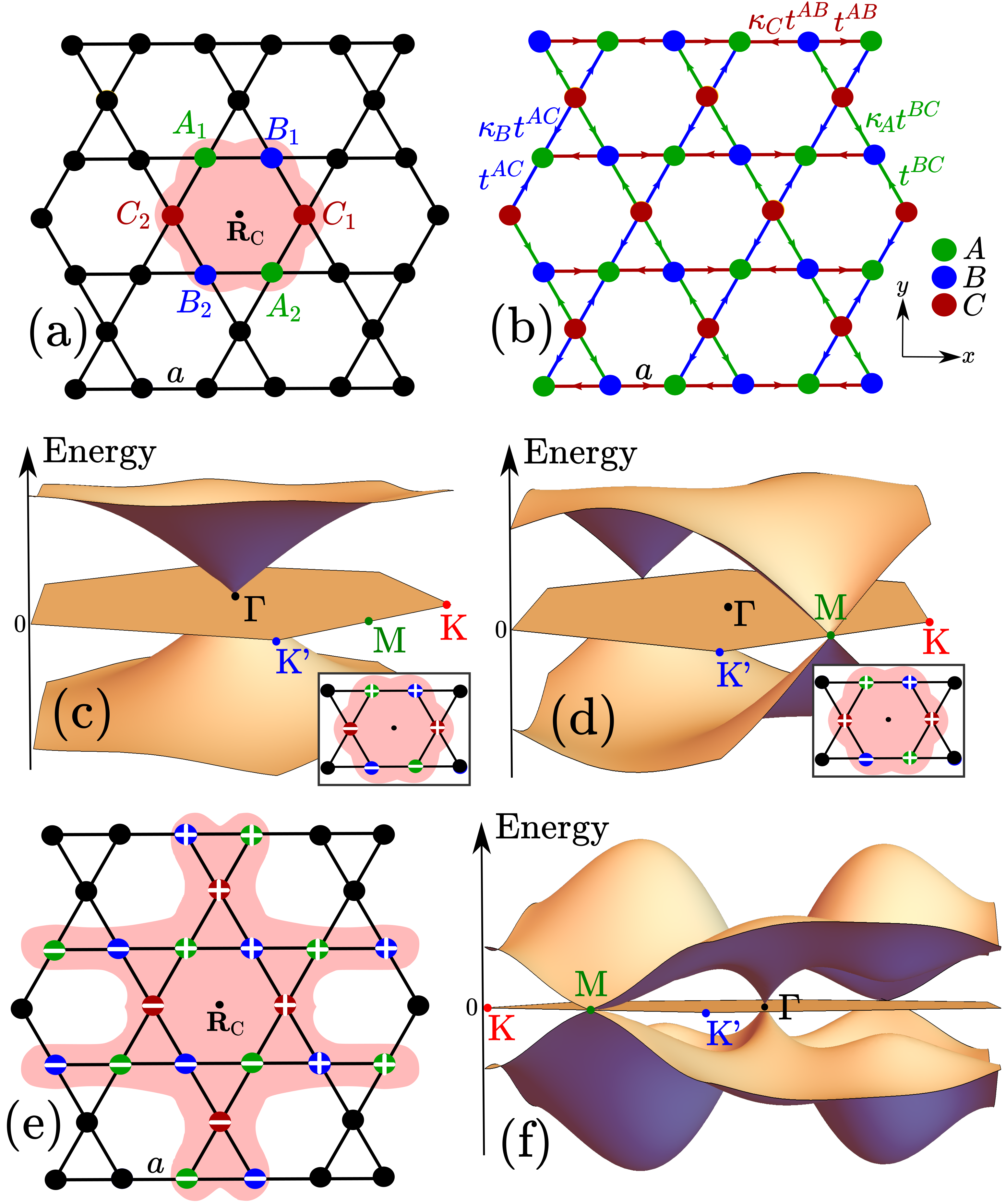}
	\caption{(a) A CLS on the Kagome lattice that is type-I CP if the amplitudes are properly correlated. (b) Tight-binding model built from the CLS. The corresponding band structure with $w_{\alpha_1}=1$ is shown in (c) for $\kappa_\alpha=-1$ and in (d) for $\kappa_A=-\kappa_B=\kappa_C=1$. (e) A more extended CP$_I$-CLS on the Kagome lattice, with $w_{\alpha_i}=\pm1$ as shown. (f) Band structure for the corresponding TB model.}
	\label{fig:linearexsconj}
\end{figure}

Type-I CP-CLSs cannot be found on all $N=3$ lattices, due to the above-mentioned constraints. For example, a CP$_I$-CLS cannot exist on the dice lattice, but it can exist on the Kagome lattice. As an example, consider the CLS shown in Fig. \ref{fig:linearexsconj}(a), where we take the amplitudes $w_{\alpha_1}$ to be real for concreteness.
In order for the CLS to qualify as CP$_I$, the three amplitudes $w_{\alpha_2}$ have to be chosen as $w_{\alpha_2}=\kappa_\alpha w_{\alpha_1}$, with $\kappa_A\kappa_B\kappa_C=-1$. The corresponding BCLS (\ref{conjBCLS}) is given by
\begin{equation}
	\begin{aligned}
f_A&=w_{A_1}\left(e^{-ik_-}+\kappa_Ae^{ik_-}\right),\\
f_B&=w_{B_1}\left(e^{ik_+}+\kappa_Be^{-ik_+}\right),\\ 
f_C&=w_{C_1}\left(e^{ik_x}+\kappa_Ce^{-ik_x}\right).
\end{aligned}
\label{conjf0}
\end{equation}
Inserting into Eq. (\ref{linim}), we find a flat-band TB model
\begin{equation}
H_\mathbf{k}=\begin{bmatrix}
0 & H_{AB} & H_{AC}\\
H_{AB}^* & 0 & H_{BC}\\
H_{AC}^* & H_{BC}^* & 0
\end{bmatrix},
\label{f32ham}
\end{equation}
where
\begin{equation}
\begin{aligned}
H_{AB}&=t^{AB}\left(e^{ik_x}+\kappa_Ce^{-ik_x}\right),\\
H_{AC}&=t^{AC}\left(e^{i k_+}+\kappa_Be^{-i k_+}\right),\\
H_{BC}&=t^{BC}\left(e^{-i k_-}+\kappa_Ae^{i k_-}\right).
\end{aligned}
\end{equation}
 Here, the hopping parameters are $t^{AB}\equiv-w_{C_1}$, $t^{AC}\equiv w_{B_1}$ and $t^{BC}\equiv \kappa_C w_{A_1}$. This model is shown in Fig. \ref{fig:linearexsconj}(b). Again, the band structure can be designed using the CLS amplitudes. An example with a BTP at the $\Gamma$ point is shown in Fig. \ref{fig:linearexsconj}(c) -- this is equivalent to a "breathing Kagome" lattice \cite{Green_2010} -- and one with a BTP at the M point in Fig. \ref{fig:linearexsconj}(d).
 
 As a more complicated example, consider the extended CP$_I$-CLS shown in Fig. \ref{fig:linearexsconj}(e). Inserting into Eq. (\ref{linim}) leads to a flat-band TB model with linear BTPs at the $\Gamma$ and M points, as shown in Fig. \ref{fig:linearexsconj}(f).
 
Finally, an example for a CP$_I$-CLS on a 3D lattice is given in Appendix \ref{Appcube}.
 

\subsubsection*{Type-II CP-CLSs}

The fourth possibility for building an $N = 3$ linear flat-band Hamiltonian arises if the input CLS (\ref{3CLSb}) occupies all sublattices, with the following constraints: within the CLS, the orbital positions of one sublattice and a second sublattice are exchanged upon inversion (with the localization center being the center of inversion), and the CLS amplitudes are related by complex conjugation. At the same time, the third sublattice (indicated by $\tau$) has orbitals (position and amplitude) that are correlated in a symmetric fashion with respect to the localization center. We may call such a CLS \emph{type-II CP (CP$_{II}$)}.
More precisely, any CP$_{II}$-CLS corresponds to a \emph{CP$_{II}$-BCLS}
\begin{equation}
	\begin{aligned}
		\ket{f_\mathbf{k}}&=|q_{\tau,\mathbf{k}}^{ABC}\rangle,\\
		|q_{A,\mathbf{k}}^{ABC}\rangle&\equiv(f_A,-f_C^*,f_C)^T,\\
		|q_{B,\mathbf{k}}^{ABC}\rangle&\equiv(f_A,f_B,-f_A^*)^T,\\ 
		|q_{C,\mathbf{k}}^{ABC}\rangle&\equiv(-f_B^*,f_B,f_C)^T,
	\end{aligned}
	\label{conjuBCLS}
\end{equation}
where the component $f_\tau$ is real. Consequently, $f_Af_Bf_C=(f_Af_Bf_C)^*$.

It is easy to see that the matrix \smash{$Q_{\tau,\mathbf{k}}^{ABC}$}, with
\begin{equation}
	\begin{aligned}
		Q_{A,\mathbf{k}}^{ABC}&\equiv\begin{bmatrix}
			0&f_C&f_C^*\\
			f_C^*&f_A&0\\
			f_C & 0 & -f_A
		\end{bmatrix},\\
		Q_{B,\mathbf{k}}^{ABC}&\equiv\begin{bmatrix}
			-f_B & f_A & 0\\
			f_A^* & 0 & f_A\\
			 0 & f_A^* & f_B
		\end{bmatrix},\\	Q_{C,\mathbf{k}}^{ABC}&\equiv\begin{bmatrix}
			f_C&0&f_B^*\\
			0&-f_C&f_B\\
			f_B & f_B^* & 0
		\end{bmatrix},
	\end{aligned}
\end{equation}
vanishes on \smash{$|q_{\tau,\mathbf{k}}^{ABC}\rangle$}.
Thus, any CP$_{II}$-CLS can be used to build a linear flat-band Bloch Hamiltonian
\begin{equation}
	H_\mathbf{k}=\lambda_\mathbf{k}^{ABC}Q_{\tau,\mathbf{k}}^{ABC}.
	\label{HABCcon}
\end{equation} 
This Hamiltonian is in fact a \emph{pseudospin-1 Hamiltonian}, i.e. it can be written in the form $H_\mathbf{k}=h_x(\mathbf{k}) S_x+h_y(\mathbf{k})S_y+h_z(\mathbf{k})S_z$, with pseudospin-1 operators that fulfill the algebra $\comm{S_i}{S_j}=i\epsilon_{ijk}S_k$. It has a band structure
\begin{equation}
	\begin{aligned}
		\epsilon_0(\mathbf{k})&=0,\\
		\epsilon_{1,2}(\mathbf{k})&=\pm\lambda_\mathbf{k}^{ABC}\sqrt{\langle q_{\tau,\mathbf{k}}^{ABC}|q_{\tau,\mathbf{k}}^{ABC}\rangle},
	\end{aligned}
	\label{bands3}
\end{equation} 
whose particle-hole symmetric character is protected by a CP symmetry (\ref{chargecon}) with $\mathcal{C}=e^{i\pi S_y}$, with diagonal matrix elements $\mathcal{C}_{\alpha\alpha}=-\delta_{\alpha\tau}$ and off-diagonal elements $\mathcal{C}_{\alpha\beta}=1-\delta_{\alpha\tau}-\delta_{\beta\tau}$. Again \smash{$\lambda_\mathbf{k}^{ABC}=1$} in the examples below.

\begin{figure}
	\centering
	\includegraphics[width=\columnwidth]{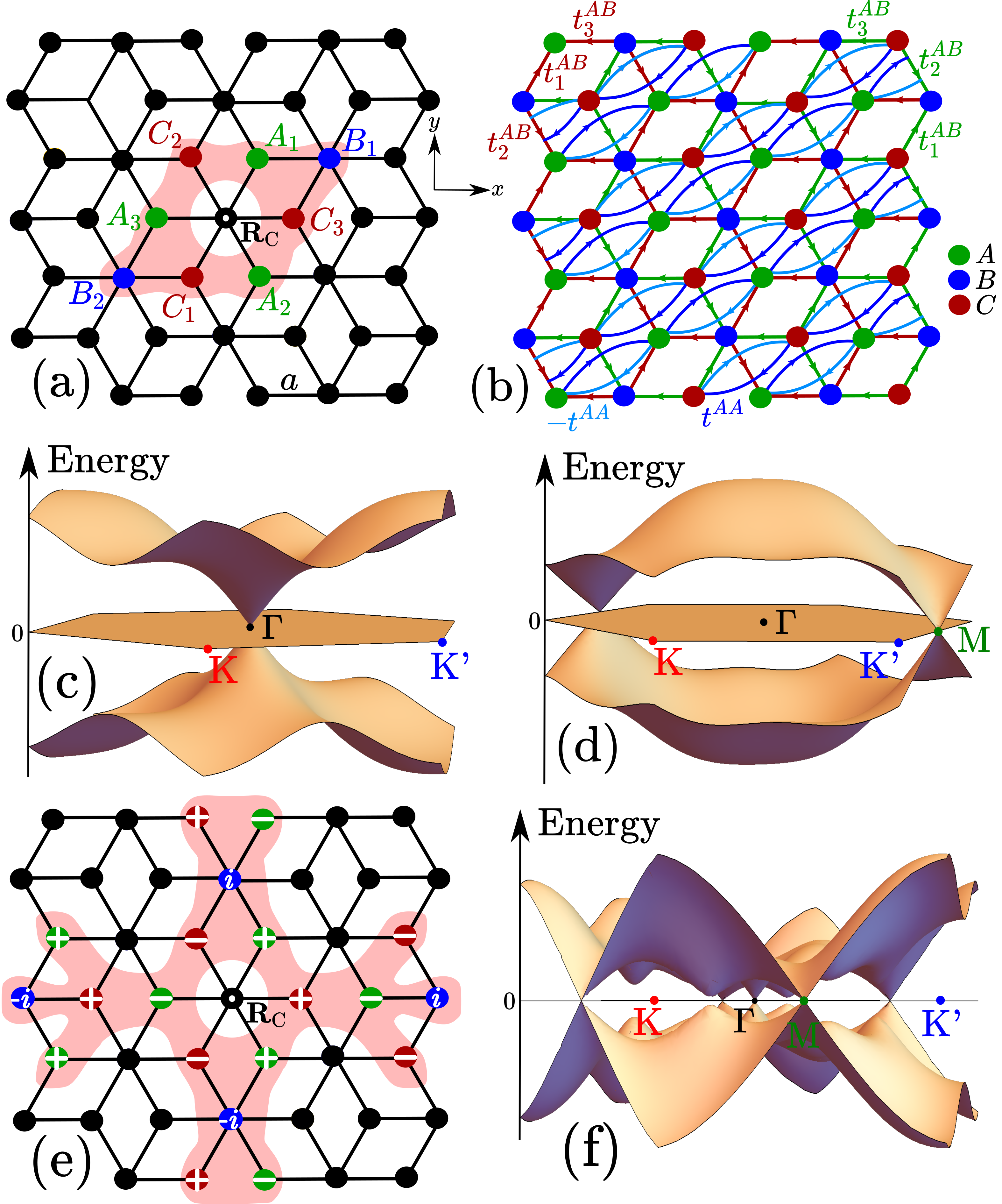}
	\caption{(a) A CLS on the dice lattice that is type-II CP if the amplitudes are properly correlated. (b) Tight-binding model built from the CLS. The corresponding band structure is shown in (c) for $w_{A_1}=w_{A_2}=-w_{A_3}/2=1$, $w_{B_1}=i$ and in (d) for $w_{A_1}=w_{A_2}=w_{A_3}/2=1$, $w_{B_1}=i$. (e) A more extended CP$_{II}$-CLS on the dice lattice, with $w_{A_i}=\pm1$ and $w_{B_i}=\pm i$ as shown. (f) Band structure for the corresponding TB model.}
	\label{fig:spinexs}
\end{figure}

Type-II CP-CLSs cannot be found on all $N=3$ lattices due to the requirement of two sublattices being the conjugate of each other. For example, a CP$_{II}$-CLS cannot exist on the Kagome lattice, but it can exist on the dice lattice. As an example, consider the CLS shown in Fig. \ref{fig:spinexs}(a), where we fix $w_{B_i}$ to be imaginary for concreteness.
 In order for the CLS to qualify as CP$_{II}$, we now have to take $w_{B_2}=-w_{B_1}$ such that $\tau=B$, and $w_{C_i}=-w_{A_i}$. The corresponding BCLS (\ref{conjuBCLS}) is then given by
 \begin{equation}
 \begin{aligned}
 f_A&=w_{A_1}e^{ik_+}+w_{A_2}e^{ik_-}+w_{A_3}e^{-ik_x},\\
 f_B&=2iw_{B_1}\sin(k_x+k_+).
 \end{aligned}
 \end{equation}
Inserting into Eq. (\ref{HABCcon}), we find a flat-band TB model
\begin{equation}
H_\mathbf{k}=\begin{bmatrix}
	H_{AA} & H_{AB} & 0\\
H_{AB}^* & 0 & H_{BC}\\
0 & H_{BC}^* & H_{CC}
\end{bmatrix},
\end{equation}
where
\begin{equation}
	\begin{aligned}
H_{AA}&=-H_{CC}=2it^{AA}\sin(k_x+k_+),\\
H_{AB}&=H_{BC}=t_1^{AB}e^{ik_+}+t_2^{AB}e^{ik_-}+t_3^{AB}e^{-ik_x},
\end{aligned}
\end{equation}
with hopping parameters $t^{AA}\equiv-w_{B_1}$ and $t_i^{AB}\equiv w_{A_i}$. The corresponding real-space TB model is shown in Fig. \ref{fig:spinexs}(b), and its band structure can be designed using the CLS amplitudes. An example with a BTP at the $\Gamma$ point is shown in Fig. \ref{fig:spinexs}(c), and one with a BTP at the M point in Fig. \ref{fig:spinexs}(d). Note that BTPs at the K points could be created by taking an input CLS with six $B$ orbitals arrayed hexagonally.

As a more complicated example, consider the extended CP$_{II}$-CLS shown in Fig. \ref{fig:spinexs}(e). Inserting into Eq. (\ref{HABCcon}) leads to a flat-band TB model with linear BTPs at the $\Gamma$ and M points, as shown in Fig. \ref{fig:spinexs}(f). 

Finally, an example for a CP$_{II}$-CLS in 3D is given in Appendix \ref{Appcube}.

\subsubsection{Conditions for reasonable linear flat-band models}

As discussed above, a given CLS has to possess special properties to qualify as chiral or as CP. However, even if some given CLS is, say, type-I chiral, it is not automatically guaranteed that the Hamiltonian (\ref{HABC}) constructed from it will be \emph{reasonable}, i.e. make sense when transformed to real space [cf. condition (\ref{constr2})]. The same is true for the other three classes of CLSs.

\begin{figure}
	\centering
	\includegraphics[width=\columnwidth]{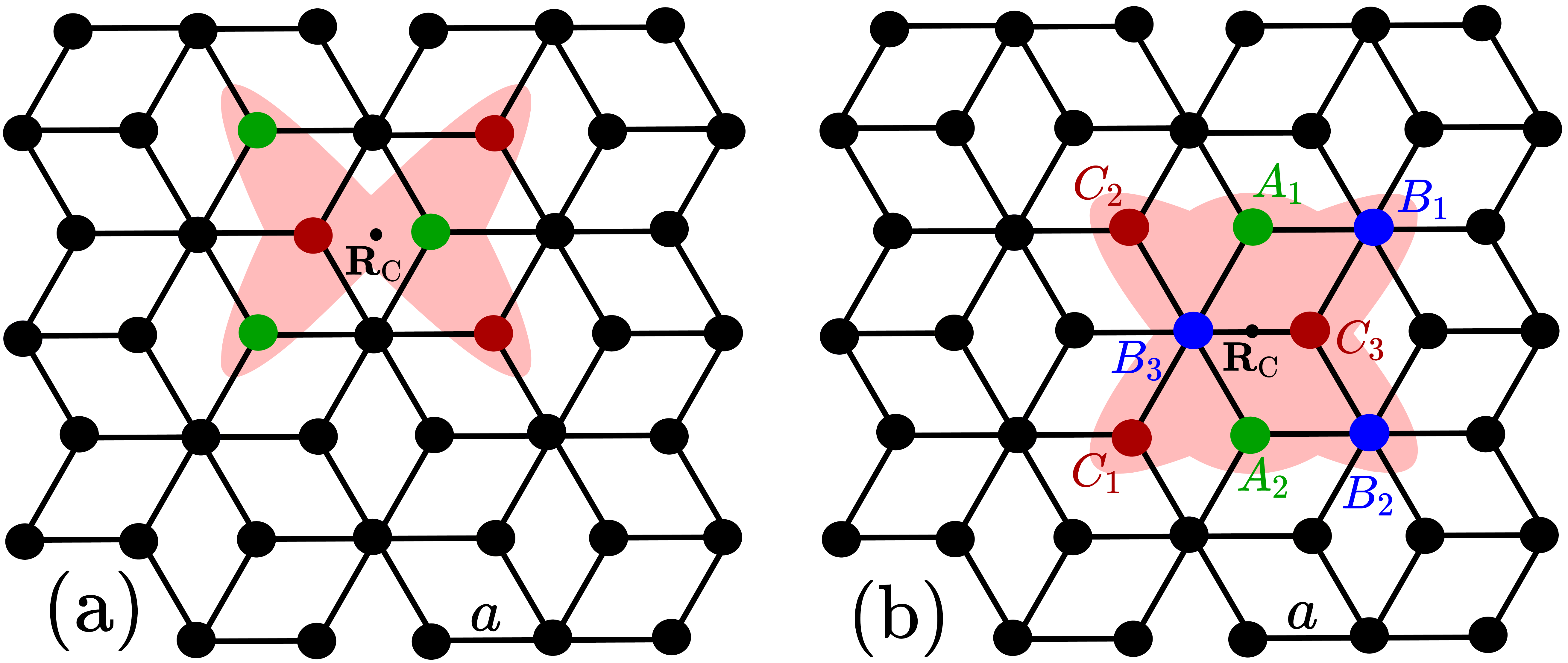}
	\caption{(a) A type-I chiral CLS on the dice lattice for which the linear Hamiltonian (\ref{HABC}) is not physically reasonable. (b) A type-II CP-CLS on the dice lattice for which the linear Hamiltonian (\ref{HABCcon}) is not physically reasonable.}
	\label{fig:counterexs}
\end{figure}

In Fig. \ref{fig:linearexs}, we have seen C$_I$-CLSs for which Eq. (\ref{HABC}) does make sense on a lattice. As a counterexample, consider the CLS shown in Fig. \ref{fig:counterexs}(a).
 It perfectly qualifies as a C$_I$-CLS with $\tau=B$, however we can easily convince ourselves that the Bloch Hamiltonian obtained by inserting its BCLS into Eq. (\ref{HABC}) does not correspond to a reasonable lattice model. 
 Similarly, in Fig. \ref{fig:spinexs}, we have seen CP$_{II}$-CLSs for which Eq. (\ref{HABCcon}) makes sense on a lattice. As a counterexample, consider the CLS shown in Fig. \ref{fig:counterexs}(b), which perfectly qualifies as a CP$_{II}$-CLS with $\tau=A$ if $w_{A_1}=w_{A_2}$ and $w_{C_i}=-w_{B_i}$, but will not lead to a reasonable real-space model when its BCLS is inserted into Eq. (\ref{HABCcon}).

There are thus two fundamental reasons that make the linear flat-band models much rarer than the quadratic models of Section \ref{quadmods}: (i) Only for certain classes of input CLSs it is possible to construct a linear matrix $H_\mathbf{k}$ that fulfills $H_\mathbf{k}\ket{f_\mathbf{k}}=0.$ (ii) Even if the input CLS belongs to such a special class, the position of the localization center and the arrangement of the sites around it need to be compatible with the underlying lattice (cf. Appendix \ref{AppA}), else the linear matrix $H_\mathbf{k}$ will not make sense in real space.


\subsection{$N$-band models}
\label{linconstr}

We now extend the notion of a linear flat-band Hamiltonian to any $N$. Thus, consider again a CLS (\ref{CLSdef}) built on a lattice with $N$ orbitals per unit cell, associated to a BCLS 
 \begin{equation}
 \ket{f_\mathbf{k}}=(f_A,f_B,f_C,...)^T.
 \label{BCLSN}
\end{equation}
The procedure that we will adopt for building a linear flat-band Hamiltonian from such a CLS is very similar in spirit to what was done in Section \ref{quadmods} for the quadratic flat-band models. There, we identified the $N=2$ quadratic Hamiltonian (\ref{2Ham}) as a \emph{building block} that allowed to construct $N\times N$ matrices \smash{$F_\mathbf{k}^{\alpha\beta}$} that vanish on $\ket{f_\mathbf{k}}$; the $N$-band quadratic Hamiltonian (\ref{mainham}) was then formed as a linear combination of these matrices.

 In exactly the same way, we can now regard the $N=3$ linear Hamiltonians (\ref{HABC}), (\ref{HABC2}), (\ref{linim}) and (\ref{HABCcon}) as \emph{building blocks} that will allow to construct $N\times N$ matrices $F_\mathbf{k}^{\alpha\beta\gamma}$ that vanish on $\ket{f_\mathbf{k}}$. An $N$-band linear flat-band Hamiltonian can then be formed as a linear combination of these matrices,
 \begin{equation}
 	H_\mathbf{k}=\sum_{\alpha,\beta>\alpha,\gamma>\beta}\lambda_\mathbf{k}^{\alpha\beta\gamma}F_\mathbf{k}^{\alpha\beta\gamma},
 	\label{mainlin}
 \end{equation} 
where \smash{$\lambda_\mathbf{k}^{\alpha\beta\gamma}$} is an arbitrary function with the periodicity of the FBZ  (\smash{$\lambda_\mathbf{k}^{\alpha\beta\gamma}=\lambda_{\alpha\beta\gamma}$} hereafter).

The Hamiltonian (\ref{mainlin}) is the second main result of this paper. It is the linear analog of the $N$-band quadratic Hamiltonian (\ref{mainham}). Several remarks about its construction and generic properties are in order.

In contrast to the quadratic Hamiltonian (\ref{mainham}), which can be built from any input CLS, a linear Hamiltonian (\ref{mainlin}) can only be built from a CLS with special properties. In particular, the CLS must be decomposable into chiral and/or CP-type \emph{sub-CLSs}
\begin{equation}
	\ket{\psi_{\alpha\beta\gamma}}=\sum_{i\in\text{CLS}}(w_{\alpha_i}\ket{\alpha_i}+w_{\beta_i}\ket{\beta_i}+w_{\gamma_i}\ket{\gamma_i}),
	\label{subCLS}
\end{equation} 
each of which involves only three sublattices $(\alpha,\beta,\gamma)$ out of the $N$ sublattices. As described in detail in Appendix \ref{Nconstr}, each sub-CLS contributes one term to the Hamiltonian (\ref{mainlin}), more precisely \smash{$F_\mathbf{k}^{\alpha\beta\gamma}=M_{\tau,\mathbf{k}}^{\alpha\beta\gamma}$}, \smash{$F_\mathbf{k}^{\alpha\beta\gamma}=N_{\tau,\mathbf{k}}^{\alpha\beta\gamma}$}, \smash{$F_\mathbf{k}^{\alpha\beta\gamma}=O_\mathbf{k}^{\alpha\beta\gamma}$}, or \smash{$F_\mathbf{k}^{\alpha\beta\gamma}=Q_{\tau,\mathbf{k}}^{\alpha\beta\gamma}$}, if $\ket{\psi_{\alpha\beta\gamma}}$ is a type-I chiral, type-II chiral, type-I CP or type-II CP sub-CLS, respectively. 

The Hamiltonian (\ref{mainlin}) consists of only one term if $N=3$; in this case, it is tunable only by the input CLS, just like the $N=2$ quadratic models. 
In contrast, the $N\geq4$ linear models obtained from Eq. (\ref{mainlin}) are, just like the $N\geq3$ quadratic models, tunable by the input CLS \emph{and} the parameters $\lambda_{\alpha\beta\gamma}$ that arise from the linear combination.

If only C$_I$-type matrices $M_{\tau,\mathbf{k}}^{\alpha\beta\gamma}$ contribute to the Hamiltonian (\ref{mainlin}), it exhibits an effective (global) chiral symmetry (\ref{chiral}) with a unitary matrix $\mathcal{S}=\text{diag}(\sigma_A,\sigma_B,\sigma_C,...)$, $\mathcal{S}^2=\mathbf{1}_N$, where $\sigma_\alpha=1$ for all sublattices occupied by the CLS and $\sigma_\alpha=-1$ for all unoccupied sublattices. Similarly, if only CP$_I$-type matrices $O_\mathbf{k}^{\alpha\beta\gamma}$ contribute, $H_\mathbf{k}$ exhibits an effective (global) CP symmetry (\ref{chargecon}) with a unitary matrix $\mathcal{C}=\text{diag}(\kappa_A,\kappa_B,\kappa_C,...)$, $\mathcal{C}^2=\mathbf{1}_N$, where $\kappa_\alpha=1$ if $f_\alpha$ is real and $\kappa_\alpha=-1$ if $f_\alpha$ is imaginary. 
In the presence of either chiral or CP symmetry, the band structure of the Hamiltonian (\ref{mainlin}) takes a simple form for $N=4$ and $N=5$, respectively, due to the simple structure of the characteristic polynomial:
\begin{equation}
	\begin{aligned}
		\epsilon_n(\mathbf{k})&=\left\{0,0,\pm \frac{C_2}{2}\right\},\\
		\epsilon_n(\mathbf{k})&=\left\{0,\pm\sqrt{\frac{C_2}{4}\pm\frac{1}{2}\sqrt{C_4-\frac{C_2^2}{4}}}\right\},
	\end{aligned}
	\label{Symbands}
\end{equation}
where again $C_n\equiv\Tr(H_\mathbf{k}^n)$.
 In contrast, if only C$_{II}$-type matrices \smash{$N_{\tau,\mathbf{k}}^{\alpha\beta\gamma}$}, or only CP$_{II}$-type matrices \smash{$Q_{\tau,\mathbf{k}}^{\alpha\beta\gamma}$}, or matrices of different types contribute to the Hamiltonian (\ref{mainlin}), no chiral or CP symmetry will in general be present. In this case, there is no simple closed-form expression for the dispersive bands, but the existence of a flat band is always guaranteed. 

Finally, just like the quadratic Hamiltonian (\ref{mainham}), the linear Hamiltonian (\ref{mainlin}) vanishes at $\mathbf{k}_0$ whenever $\ket{f_{\mathbf{k}_0}}=0$. As a consequence, there will be an $N$-fold (singular) BTP at $\mathbf{k}_0$. More generally, in contrast to the quadratic BTPs of Section \ref{quadmods}, the linear BTPs encountered here can of course not be of any degeneracy, as they must involve an odd number of bands.

\begin{figure}
	\centering
	\includegraphics[width=\columnwidth]{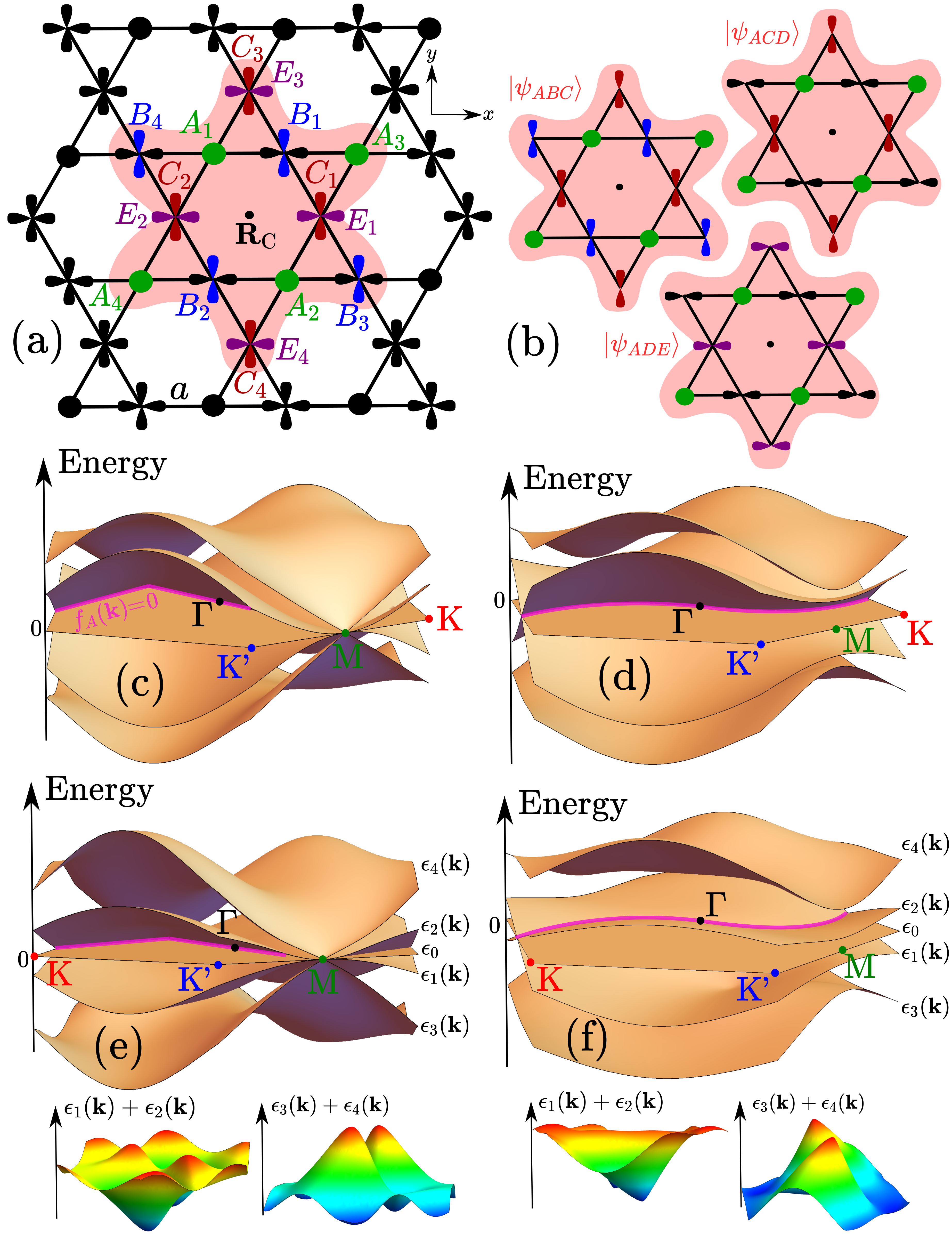}
	\caption{(a) CLS on an $N=5$ Kagome lattice. The $D$ orbitals are at the same sites as the $B$ orbitals but are unoccupied, as indicated by the black color. (b) Decomposition into three sub-CLSs. (c) \& (d) Band structure of the corresponding TB model (\ref{N5conj}), with $\lambda_{ABC}=\lambda_{ADE}=1$, $\lambda_{ACD}=0$, for (c) $w_{\alpha_1}=w_{\alpha_3}=1$ and (d) $w_{\alpha_1}=2w_{\alpha_3}=2$, where $\alpha=A,B,C,E$. (e) \& (f) Same as (c) \& (d) but with $\lambda_{ACD}=1$. The insets show the deviation from particle-hole symmetry of the bands.}
	\label{fig:conjchiral}
\end{figure}

As an example that well illustrates many of these generic features, consider the CLS on the multiorbital ($N=5$) Kagome lattice shown in Fig. \ref{fig:conjchiral}(a).
We take the orbitals on sublattices $\alpha=A,B,C$ to be occupied antisymmetrically as $w_{\alpha_2}=-w_{\alpha_1}$, $w_{\alpha_4}=-w_{\alpha_3}$, sublattice $D$ to be unoccupied, sublattice $E$ to be occupied symmetrically as $w_{E_2}=w_{E_1}$, $w_{E_4}=w_{E_3}$, and we will further assume $w_{\alpha_i}\in\mathbb{R}$. The corresponding BCLS then takes the form $\ket{f_\mathbf{k}}=(f_A,f_B,f_C,0,f_E)^T$, where
\begin{equation}
\begin{aligned}
f_A&=-2i[w_{A_1}\sin k_--w_{A_3}\sin(k_x+k_+)],\\
f_B&=-2i[w_{B_1}\sin k_+-w_{B_3}\sin(k_x+k_-)],\\
f_C&=2i[w_{C_1}\sin k_x+w_{C_3}\text{sin}(\sqrt{3}k_y)],\\
f_E&=2[w_{E_1}\cos k_x+w_{E_3}\text{cos}(\sqrt{3}k_y)].
\end{aligned}
\label{BCLSex5}
\end{equation}
This CLS can be decomposed into three sub-CLSs, namely a type-I CP sub-CLS $\ket{\psi_{ABC}}$ and two type-I chiral sub-CLSs $\ket{\psi_{ACD}}$ and $\ket{\psi_{ADE}}$, as shown in Fig. \ref{fig:conjchiral}(b) (see Appendix \ref{Nconstr} for details). 
Accordingly, from this CLS, we can build an $N=5$ linear flat-band Hamiltonian
\begin{equation}
H_\mathbf{k}=\lambda_{ABC}F^{ABC}_\mathbf{k}+\lambda_{ACD}F^{ACD}_\mathbf{k}+\lambda_{ADE}F^{ADE}_\mathbf{k},
\label{N5conj}
\end{equation}
where
\[
F^{ABC}_\mathbf{k}=O^{ABC}_\mathbf{k}\equiv\begin{bmatrix}
0 & -f_C & f_B & 0 & 0\\
f_C & 0 & -f_A & 0 & 0\\
-f_B & f_A & 0 & 0 & 0\\
0 & 0 & 0 & 0 & 0\\
0 & 0 & 0 & 0 & 0
\end{bmatrix},
\]
\[
F^{ACD}_\mathbf{k}=M^{ACD}_{D,\mathbf{k}}\equiv\begin{bmatrix}
	0 & 0 & 0 & -f_C^* & 0\\
	0 & 0 & 0 & 0 & 0\\
	0 & 0 & 0 & f_A^* & 0\\
	-f_C & 0 & f_A & 0 & 0\\
	0 & 0 & 0 & 0 & 0
\end{bmatrix},\]
and
\[
F^{ADE}_\mathbf{k}=M^{ADE}_{D,\mathbf{k}}\equiv\begin{bmatrix}
	0 & 0 & 0 & -f_E^* & 0\\
	0 & 0 & 0 & 0 & 0\\
	0 & 0 & 0 & 0 & 0\\
	-f_E & 0 & 0 & 0 & f_A\\
	0 & 0 & 0 & -f_A & 0
\end{bmatrix}.\]
Two interesting cases (i.e. where no band is decoupled) can be identified, namely $\lambda_{ABC}, \lambda_{ADE}\neq0$ with either $\lambda_{ACD}=0$ or $\lambda_{ACD}\neq0$. In the former case, the Hamiltonian has an effective CP symmetry with $\mathcal{C}=\text{diag}(1,1,1,-1,-1)$, such that the energy spectrum is of the form (\ref{Symbands}). An example for this is shown in Fig. \ref{fig:conjchiral}(c), where fivefold BTPs appear at the M points. They are connected by a threefold nodal line at $f_A=0$. If some imbalance is introduced in the CLS amplitudes, the fivefold degeneracy is reduced to a threefold one and the nodal line is deformed, see Fig. \ref{fig:conjchiral}(d).
In the latter case, $\lambda_{ACD}\neq0$, the CP symmetry is broken and the spectrum is no longer particle-hole symmetric. Examples for this are shown in Fig. \ref{fig:conjchiral}(e)\&(f), where again a nodal line appears at $f_A=0$.

\begin{figure}
	\centering
	\includegraphics[width=\columnwidth]{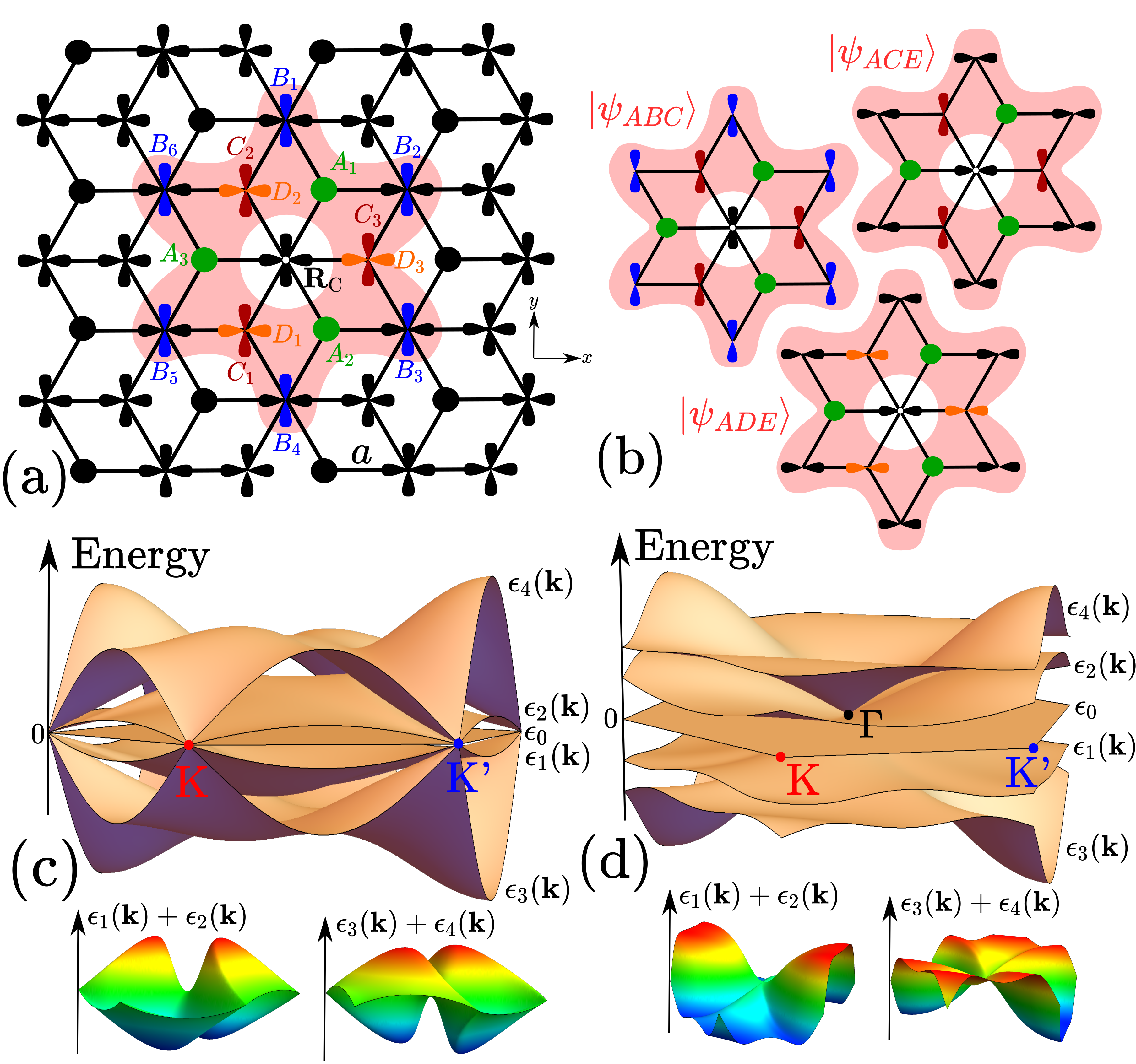}
	\caption{(a) CLS on an $N=5$ dice lattice. The $E$ orbitals are at the same sites as the $B$ orbitals but are unoccupied. (b) Decomposition into three sub-CLSs. (c) \& (d) Band structure of the corresponding TB model (\ref{N5spin}), with $\lambda_{ABC}=\lambda_{ACE}=\lambda_{ADE}=1$ and $w_{B_{1,4}}=-2w_{B_{2,3,5,6}}=-2$, for (c) $w_{A_i}=w_{D_i}=1$ and (d) $w_{A_{1,2}}=w_{D_{1,2}}=-w_{A_3}/2=-w_{D_3}/2=1$. The insets show the deviation from particle-hole symmetry of the bands.}
	\label{fig:spinchiral}
\end{figure}

As a second example, consider the CLS on the multiorbital ($N=5$) dice lattice shown in Fig. \ref{fig:spinchiral}(a).
We take the orbitals on sublattice $B$ to be occupied symmetrically as $w_{B_4}=w_{B_1}$, $w_{B_5}=w_{B_2}$, $w_{B_6}=w_{B_3}$, sublattice $C$ to be occupied oppositely to sublattice $A$, $w_{C_i}=-w_{A_i}$, sublattice $E$ to be unoccupied, and we will further assume $w_{\alpha_i}\in\mathbb{R}$. The corresponding BCLS takes the form $\ket{f_\mathbf{k}}=(f_A,f_B,f_C,f_D,0)^T$, where
\begin{equation}
	\begin{aligned}
		f_A&=w_{A_1}e^{ik_+}+w_{A_2}e^{ik_-}+w_{A_3}e^{-ik_x},\\
		f_B&=2[w_{B_1}\text{cos}(\sqrt{3}k_y)+w_{B_2}\text{cos}(k_x+k_+)\\
		&\hspace{.5cm}+w_{B_3}\text{cos}(k_x+k_-)],\\
		f_C&=-f_A^*,\\
		f_D&=w_{D_1}e^{-ik_+}+w_{D_2}e^{-ik_-}+w_{D_3}e^{ik_x}.
	\end{aligned}
	\label{BCLSexx5}
\end{equation}
This CLS can be decomposed into three sub-CLSs, namely a type-II CP sub-CLS $\ket{\psi_{ABC}}$ and two type-I chiral sub-CLSs $\ket{\psi_{ACE}}$ and $\ket{\psi_{ADE}}$, as shown in Fig. \ref{fig:spinchiral}(b) (see Appendix \ref{Nconstr} for details). Accordingly, from this CLS, we can again build an $N=5$ linear flat-band Hamiltonian as a linear combination of three matrices,
\begin{equation}
	H_\mathbf{k}=\lambda_{ABC}F^{ABC}_\mathbf{k}+\lambda_{ACE}F^{ACE}_\mathbf{k}+\lambda_{ADE}F^{ADE}_\mathbf{k},
	\label{N5spin}
\end{equation}
where now
\[
		F^{ABC}_\mathbf{k}=Q^{ABC}_{B,\mathbf{k}}\equiv\begin{bmatrix}
			-f_B & f_A & 0 & 0 & 0\\
			f_A^* & 0 & f_A & 0 & 0\\
			0 & f_A^* & f_B & 0 & 0\\
			0 & 0 & 0 & 0 & 0\\
			0 & 0 & 0 & 0 & 0
		\end{bmatrix},\]
	\[
		F^{ACE}_\mathbf{k}=M^{ACE}_{E,\mathbf{k}}\equiv\begin{bmatrix}
			0 & 0 & 0 & 0 & -f_C^*\\
			0 & 0 & 0 & 0 & 0\\
			0 & 0 & 0 & 0 & f_A^*\\
		    0 & 0 & 0 & 0 & 0\\
			-f_C & 0 & f_A & 0 & 0		
		\end{bmatrix},\]
	and
	\[
		F^{ADE}_\mathbf{k}=M^{ADE}_{E,\mathbf{k}}\equiv\begin{bmatrix}
			0 & 0 & 0 & 0 & -f_D^*\\
			0 & 0 & 0 & 0 & 0\\
			0 & 0 & 0 & 0 & 0\\
			0 & 0 & 0 & 0 & f_A^*\\
			-f_D & 0 & 0 & f_A & 0
		\end{bmatrix}.\]
For all interesting cases (i.e. where no band is decoupled) the spectrum is not particle-hole symmetric, due to the nontrivial mixing of chiral and CP-type building blocks. In Fig. \ref{fig:spinchiral}(c)\&(d), the spectrum is shown for different amplitudes of the input CLS. In particular, fivefold BTPs can be created at the K points or at the $\Gamma$ point. Of course, the bands are also tunable by the three parameters $\lambda_{\alpha\beta\gamma}$.

\section{Interpolating between linear and quadratic band touching}
\label{mixedmods}

From the previous discussion we know that all CLSs allow to build quadratic flat-band models, while only special CLSs allow to build linear ones. In other words, each such special CLS can be used to create both a linear \emph{and} a quadratic flat-band model; both models will have a zero-energy flat band and an \emph{identical} flat-band eigenstate. In this way, we can construct tight-binding models that allow to \emph{interpolate smoothly} between multifold linear and quadratic BTPs, without moving the BTP or destroying the flat band.

\begin{figure}
	\centering
	\includegraphics[width=\columnwidth]{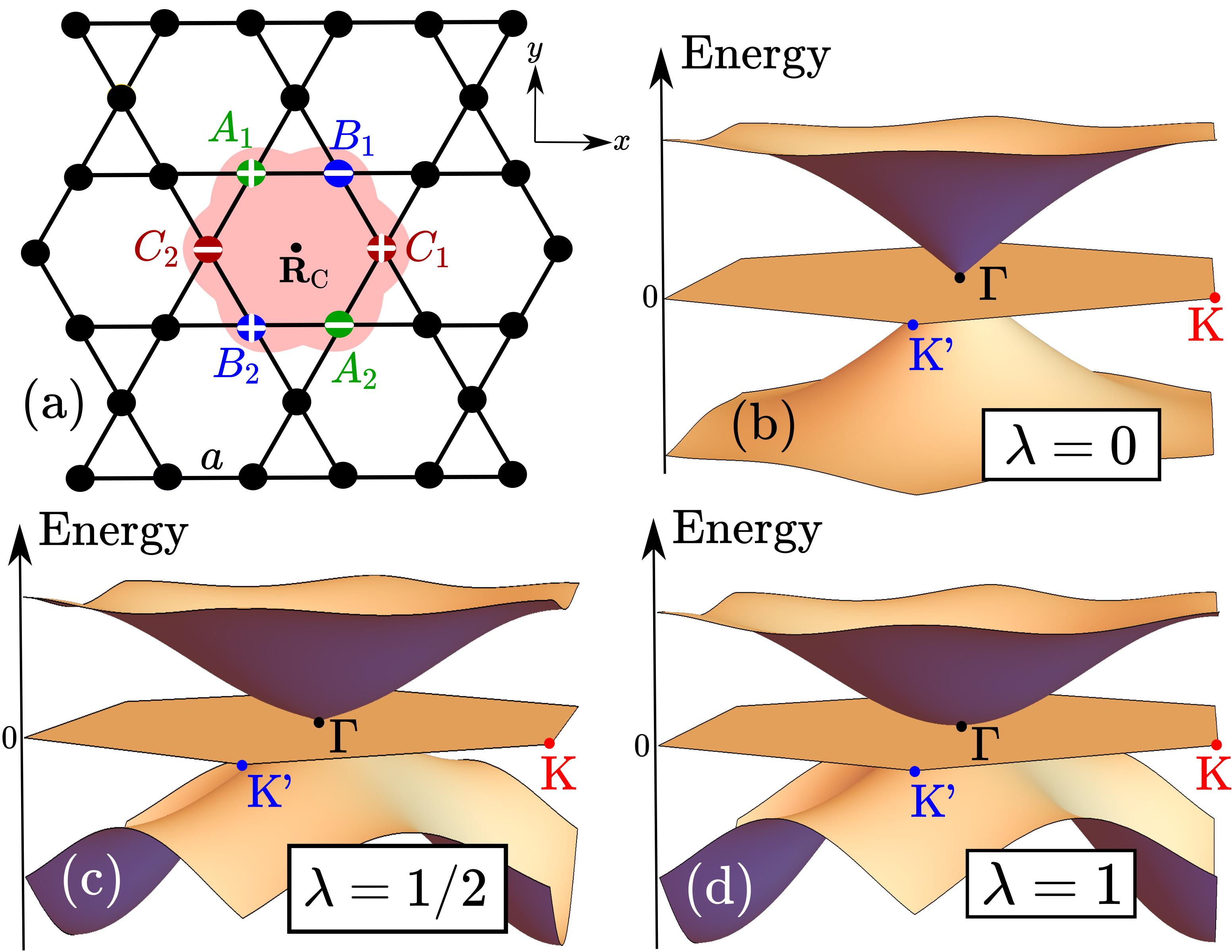}
	\caption{(a) A CLS on the Kagome lattice. (b)--(d) Band structure of the mixed linear-quadratic TB model (\ref{interpol}) built from it, for different values of $\lambda$.}
	\label{fig:interpolate}
\end{figure}

As a simple example, consider again a six-site CLS on the Kagome lattice, where we now fix $w_{A_1}=-w_{B_1}=w_{C_1}=-w_{A_2}=w_{B_2}=-w_{C_2}=1$ for simplicity, as shown in Fig. \ref{fig:interpolate}(a).
This CLS is clearly of CP$_I$-type, with BCLS $\ket{f_\mathbf{k}}=(f_A,f_B,f_C)^T=-2i(\sin k_-,\sin k_+,-\sin k_x)^T$, and can thus be used to create a linear Bloch Hamiltonian $H_\text{lin}=O_\mathbf{k}^{ABC}$, cf. Eq. (\ref{linim}). But it can also be used to create a quadratic Bloch Hamiltonian $H_\text{quad}=\lambda_{AB}F_\mathbf{k}^{AB}+\lambda_{AC}F_\mathbf{k}^{AC}+\lambda_{BC}F_\mathbf{k}^{BC}$, cf. Eq. (\ref{3Ham}). We may now interpolate between the two,
\begin{equation}
H_\mathbf{k}=(1-\lambda) H_\text{lin}+\lambda H_\text{quad},
\label{interpol}
\end{equation}
which gives rise to a tight-binding model that can be smoothly tuned from linear to quadratic by the parameter $\lambda\in[0,1]$, \emph{without} the flat band and its eigenstate being affected in any way. This is illustrated in Fig. \ref{fig:interpolate}(b)--(d).

In the same way, all the linear BTPs encountered in Section \ref{linmods} can be smoothly tuned to become quadratic, simply by interpolating the respective linear flat-band model with the quadratic model (\ref{mainham}) built from the same CLS.

\section{Conclusions}
\label{concl}

It has been known for a long time that, for any given flat-band tight-binding (TB) model $H$ on a periodic lattice, it is possible to find a macroscopic number of compact localized states (CLSs) as eigenstates of $H$.

Conversely, in the present work, we have shown that for \emph{any given (input) CLS} \smash{$|\Psi_\text{CLS}^{\mathbf{R}_\text{C}}\rangle$} on any periodic lattice, it is possible to engineer one or several families of TB Hamiltonians $H$ for which this CLS and all its translated copies are eigenstates. We have provided a precise procedure to find such TB Hamiltonians, which are \emph{flat-band models} by construction. Since infinitely many different CLSs can be constructed on any lattice, this represents an efficient flat-band construction scheme: it yields infinitely many flat-band TB models on any periodic lattice.
  
This procedure is most conveniently carried out in reciprocal space, as explained in Section \ref{latticesec}. First, we exploited the fact that there is a one-to-one correspondence between any given CLS \smash{$|\Psi_\text{CLS}^{\mathbf{R}_\text{C}}\rangle$} and its Fourier transform, i.e. the Bloch state (BCLS) $\ket{f_\mathbf{k}}$. Flat-band Bloch Hamiltonians $H_\mathbf{k}$ can then conveniently be engineered as a function of the BCLS. Once $H_\mathbf{k}$ is determined, the real-space model $H$ is straightforwardly obtained.

For any arbitrarily designed input CLS, it is always possible to construct a flat-band model $H_\mathbf{k}$ as a \emph{quadratic} function of the BCLS, as discussed in Section \ref{quadmods}. This generic quadratic Hamiltonian, which works for any number of bands $N\geq2$, is provided in Eq. (\ref{mainham}) and gives rise to TB models with quadratic multifold BTPs at the flat-band energy. In contrast, if the input CLS has certain special properties, it can be used not only to build quadratic models, but also to construct a Bloch Hamiltonian $H_\mathbf{k}$ as a \emph{linear} function of the BCLS, as discussed in Section \ref{linmods}. This generic linear Hamiltonian, which works for any number of bands $N\geq3$, is given in Eq. (\ref{mainlin}) and provides TB models with linear multifold BTPs at the flat-band energy.
  
The generic quadratic Hamiltonian (\ref{mainham}) is based on $N=2$ building blocks of the form (\ref{2Ham}). Similarly, the generic linear Hamiltonian (\ref{mainlin}) is based on $N=3$ building blocks of the form (\ref{HABC}), (\ref{HABC2}), (\ref{linim}) and (\ref{HABCcon}). Due to this superposition principle, our models can not only be designed by choice of the input CLS, but are also widely \emph{tunable} by a set of completely free parameters. For the quadratic flat-band models, there are exactly $N_f=\binom{N}{2}$ free parameters $\lambda_{\alpha\beta}$, independently of the input CLS. In contrast, for the linear models, the precise number $N_f\leq\binom{N}{3}$ of free parameters $\lambda_{\alpha\beta\gamma}$ depends on the input CLS. Importantly, for both classes of models, the flat band and its eigenstate are \emph{independent} of the free parameters, while the dispersive bands and their eigenstates depend on them.

Since our flat-band models are tunable by these two knobs, the input CLS and the free parameters, they allow for extensive control over the existence, degeneracy, location and (non-)singularity of multifold BTPs. This is perhaps most strikingly illustrated by the discussion of Section \ref{mixedmods}: from any CLS that leads to a linear flat-band model, we can also construct TB models with BTPs whose low-energy dispersion can be smoothly interpolated from linear to quadratic. From a general perspective, this control over BTPs originates from the fact that our approach classifies flat bands and their eigenstates according to their (non-)singularity, similar in spirit to Ref. \cite{Rhim_2019}. This is different from other flat-band classification methods, for example those based on the shape and size of the CLS \cite{Maimaiti_2019,Maimaiti_2021}.
   
In closing this work, let us mention some perspectives. First, while our method allows to build infinitely many flat-band TB models, and while it captures most models described in the literature, it can be generalized in several ways. For example, a more general version of the quadratic Hamiltonian (\ref{mainham}) reads \smash{$H_\mathbf{k}=\sum_{\alpha,\beta>\alpha}\sum_{\gamma,\delta>\gamma}\lambda_{\alpha\beta\gamma\delta}|f_\mathbf{k}^{\alpha\beta}\rangle\langle f_\mathbf{k}^{\gamma\delta}|$}, where $\lambda_{\alpha\beta\gamma\delta}^*=\lambda_{\gamma\delta\alpha\beta}$. Similarly, one may extend the linear class of flat-band models by using $4\times4$ instead of $3\times 3$ matrices as building blocks, which might allow for designing Hamiltonians with a CP symmetry such that $\mathcal{C}^2=-\mathbf{1}_N$.
Moreover, it is clear that not all flat-band models are governed by Bloch Hamiltonians that are simply linear or quadratic functions of the flat-band eigenstate $\ket{f_{\mathbf{k}}}$. In fact, it is possible to define matrices $H_\mathbf{k}$ whose elements are higher-order polynomials (e.g. of degree three or four) in the components $f_\alpha$, and that still verify $H_\mathbf{k}\ket{f_{\mathbf{k}}}=0$. Furthermore, some flat-band Hamiltonians may actually be viewed as \emph{infinite-order} polynomials of the BCLS's components. As a prominent example for this, consider the ordinary nearest-neighbor TB model on the ($N = 3$) Kagome lattice (see Appendix \ref{Appkagopyro}). At the $\Gamma$ point, its flat-band eigenstate vanishes, $\ket{f_{\mathbf{k}=0}}=0$, and the band structure exhibits a twofold BTP. A situation of this kind cannot be captured by our Hamiltonians (\ref{mainham}) or (\ref{mainlin}), since they locally vanish whenever $\ket{f_\mathbf{k}}$ vanishes, forcing all the dispersive bands to touch the flat band. Instead, it appears that the Kagome Hamiltonian can be written in terms of simple rational functions (infinite-order polynomials) of the BCLS's components, but this property relies on very specific identities that relate the three components of the BCLS. In view of this discussion, it would be rewarding to establish all possible relations between $H_\mathbf{k}$ and $\ket{f_\mathbf{k}}$, involving both finite- and infinite-order polynomials.
   
  Second, we have only used strictly localized states as an input for the Hamiltonians (\ref{mainham}) and (\ref{mainlin}), in order to obtain short-range TB models with an exactly flat band. This implies that all flat bands presented in this work have a first Chern number $\mathcal{C}_1=0$ \cite{Chen_2014}. It might be worthwhile to use our construction scheme assuming that flat-band eigenstates
   could now be represented by power-law localized states instead of CLSs -- in particular with regard to a possible topological character of the flat bands.
   
   Third, the flat-band construction scheme discussed here exclusively for lattice models can be immediately applied to low-energy (continuum) models as well. Two cases may be distinguished in this regard. On the one hand, for any flat-band lattice model built from some given CLS, the low-energy theory around some point in the FBZ can be obtained simply by linearizing the BCLS components $f_\alpha(\mathbf{k})$, which will \emph{preserve} the flat band by construction. On the other hand, one can also make a general ansatz $f_\alpha(\mathbf{k})=\mathbf{v}_\alpha\cdot\mathbf{k}$ without having in mind a concrete lattice model, and study the properties of the models obtained from the Hamiltonians (\ref{mainham}) or (\ref{mainlin}).
   
   Finally, it would be interesting to analytically investigate the quantum geometric (and possibly topological) structures associated to our multifold BTPs \cite{Graf_2021}. In particular, the chiral symmetric and CP-symmetric classes of flat-band Hamiltonians introduced in Section \ref{linmods} are expected to behave differently in this regard.
   
   \emph{Note added:} After submission of this work, a similar approach to constructing flat-band models with BTPs starting from CLSs was proposed in Ref. \cite{Hwang_2021}.

\section*{Acknowledgements}
We thank Mateo Uldemolins for fruitful discussions.

\appendix

\section{Relation between input CLSs and real-space tight-binding models}
\label{AppA}

 In order to determine the real-space tight-binding model corresponding to the Bloch Hamiltonian $H_\mathbf{k}$ built from some input CLS, we write the off-diagonal matrix elements ($\alpha\neq\beta$) as
\begin{equation}
H_{\mathbf{k},\alpha\beta}=\sum_lt_l^{\alpha\beta}e^{i\mathbf{k}\cdot\mathbf{r}_l^{\alpha\beta}},
\label{offdia}
\end{equation}
where \smash{$\mathbf{r}_l^{\alpha\beta}$} denotes the possible hopping directions from an orbital of type $\beta$ towards orbitals of type $\alpha$, and \smash{$t_l^{\alpha\beta}$} denotes the corresponding hopping parameters. 
Similarly, we write the diagonal matrix elements as 
\begin{equation}
H_{\mathbf{k},\alpha\alpha}=V_\alpha+\sum_lt_l^{\alpha\alpha}e^{i\mathbf{k}\cdot\mathbf{r}_l^{\alpha\alpha}},
\label{dia}
\end{equation}
where $V_\alpha$ is the onsite energy, \smash{$\mathbf{r}_l^{\alpha\alpha}$} denotes the possible hopping directions from an orbital of type $\alpha$ towards other orbitals of type $\alpha$, and \smash{$t_l^{\alpha\alpha}$} are the corresponding hopping parameters.
Clearly, the expressions for the hopping directions, hopping parameters and onsite energies depend on whether the Bloch Hamiltonian is quadratic or linear, as described in the following.

\begin{figure}
	\centering
	\includegraphics[width=\columnwidth]{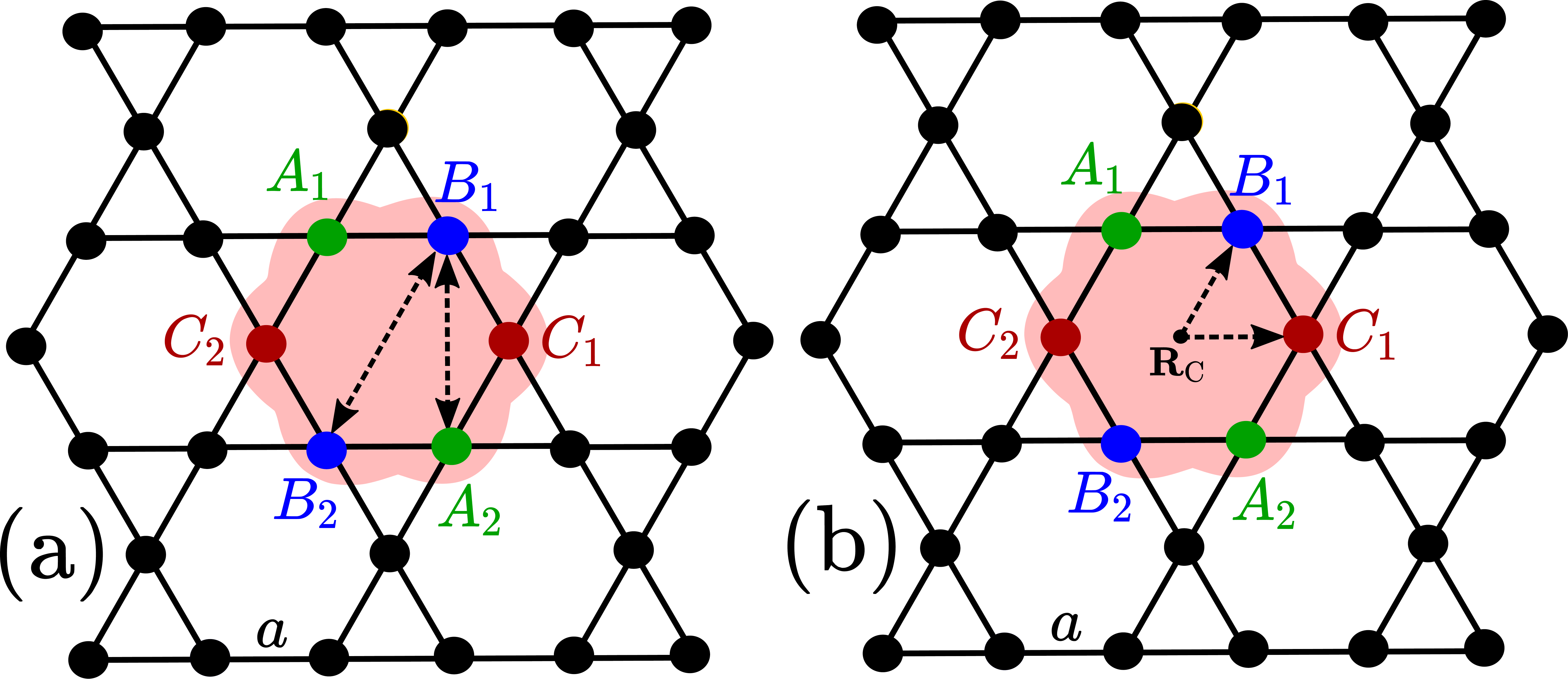}
	\caption{(a) For quadratic flat-band models, only the relative distance between each pair of orbitals within the CLS is relevant. (b) For linear flat-band models, only the position of each orbital with respect to the localization center is relevant.}
	\label{fig:quadlindiff}
\end{figure}

For the quadratic flat-band models described by Eq. (\ref{mainham}), the off-diagonal and diagonal matrix elements (\ref{matelsquad}) read
\begin{equation}
\begin{aligned}
H_{\mathbf{k},\alpha\beta}&=-\lambda_{\alpha\beta}\sum_{i,j\in\text{CLS}}w_{\alpha_i}w_{\beta_j}^*e^{i\mathbf{k}\cdot(\boldsymbol{\delta}_{\alpha_i}-\boldsymbol{\delta}_{\beta_j})},\\
H_{\mathbf{k},\alpha\alpha}&=\sum_{\beta\neq\alpha}\lambda_{\alpha\beta}\sum_{i,j\in\text{CLS}}w_{\beta_i}w_{\beta_j}^*e^{i\mathbf{k}\cdot(\boldsymbol{\delta}_{\beta_i}-\boldsymbol{\delta}_{\beta_j})},
\end{aligned}
\end{equation}
respectively, where Eq. (\ref{locdef1}) was used.
Comparing this with Eqs. (\ref{offdia}) \& (\ref{dia}), we find that the possible hopping directions $\mathbf{r}_l^{\alpha\beta}$ are given by all vectors that connect orbitals $\beta$ to orbitals $\alpha$ inside the CLS. Similarly, the hopping directions $\mathbf{r}_l^{\alpha\alpha}$ are given by all vectors that connect orbitals $\beta\neq\alpha$ to other orbitals $\beta$ inside the CLS. This is well visible in Figs. \ref{fig:CLSexample} \& \ref{fig:twobandhex}. In other words, the position of the localization center is \emph{irrelevant}, cf. Fig. \ref{fig:quadlindiff}(a).
This is exactly why any CLS will give rise to a reasonable quadratic flat-band model. The relative distance between any two orbitals is automatically compatible with the underlying lattice. For the quadratic flat-band models, the hopping parameters as well as onsite energies are functions both of the CLS amplitudes and of the parameters $\lambda_{\alpha\beta}$.

For the linear flat-band models described by Eq. (\ref{mainlin}), the exact form of the matrix elements depends on the input CLS. However, since all matrix elements are linear in $f_\alpha$, it is clear from Eq. (\ref{locdef1}) that they always contain phase factors of the form $\exp[i\mathbf{k}\cdot\boldsymbol{\delta}_{\alpha_i}]$. Generically, the hopping directions $\mathbf{r}_l^{\alpha\beta}$ are given by the positions of some orbitals of a \emph{third kind} $\gamma$ with respect to the localization center. This is well visible in Figs. \ref{fig:linearexs}--\ref{fig:spinexs}. Thus, for linear flat-band models, the position of the localization center is \emph{very important}, cf. Fig. \ref{fig:quadlindiff}(b). This is why linear flat-band models can only be built from CLSs that fulfill certain compatibility relations with the underlying lattice.
For the linear flat-band models, all onsite energies are zero. The hopping parameters are functions both of the CLS amplitudes and of the parameters $\lambda_{\alpha\beta\gamma}$.

\section{Two-band Bloch Hamiltonian with a flat band}	
\label{app2der}

To find a $2\times2$ matrix $H_\mathbf{k}$ with a (unnormalized) flat-band eigenstate $\ket{f_\mathbf{k}}$, we may first assume that $\ket{f_\mathbf{k}}$ is such an eigenstate. Evidently, this uniquely determines the corresponding eigenprojector 
\begin{equation}
	\begin{aligned}
	P_0(\mathbf{k})&\equiv\frac{\ket{f_\mathbf{k}}\bra{f_\mathbf{k}}}{\bra{f_\mathbf{k}}\ket{f_\mathbf{k}}}\\
	&=\frac{1}{|f_A|^2+|f_B|^2}\begin{bmatrix}
		|f_A|^2 & f_Af_B^*\\
		f_A^*f_B & |f_B|^2
	\end{bmatrix},
\end{aligned}
\end{equation}
but, crucially, it also fixes the orthogonal eigenprojector
\begin{equation}
		\begin{aligned}
	P_1(\mathbf{k})&=1_2-P_0(\mathbf{k})\\
	&=\frac{1}{|f_A|^2+|f_B|^2}\begin{bmatrix}
		|f_B|^2 & -f_Af_B^*\\
		-f_A^*f_B & |f_A|^2
	\end{bmatrix},
\end{aligned}
\end{equation}
This latter property is unique to the two-band case. According to the spectral theorem, the desired flat-band Hamiltonian now necessarily takes the form
\begin{equation}
		\begin{aligned}
	H_\mathbf{k}&=\epsilon_0P_0(\mathbf{k})+\epsilon_1(\mathbf{k})P_1(\mathbf{k})\\
	&=\epsilon_01_2+(\epsilon_1(\mathbf{k})-\epsilon_0)P_1(\mathbf{k}),
\end{aligned}
	\label{specdec}
\end{equation}
with the flat-band energy $\epsilon_0$. Without loss of generality, we may now take $\epsilon_0=0$, such that Eq. (\ref{specdec}) turns into Eq. (\ref{2Ham}) of the main text. In summary, all two-band models with a flat band are of the form (\ref{2Ham}).

\section{Band touching scenarios for the quadratic three-band Hamiltonian}
\label{3BTPs}

 Consider the three-band Hamiltonian (\ref{3Ham}) with band structure (\ref{spec3}). For any given point $\mathbf{k}_0$ in the FBZ, one can distinguish four possibilities for the behavior of $\ket{f_{\mathbf{k}_0}}$:

 (1) If $\ket{f_{\mathbf{k}_0}}=0$, there is a threefold (singular) BTP. 

(2) If two out of the three components of $\ket{f_{\mathbf{k}_0}}$ vanish (say $f_B(\mathbf{k}_0)=f_C(\mathbf{k}_0)=0$ for concreteness), the local Bloch Hamiltonian becomes diagonal with eigenenergies $\epsilon_0(\mathbf{k}_0)=0$, $\epsilon_1(\mathbf{k}_0)=\lambda_{AB}|f_A(\mathbf{k}_0)|^2$, $\epsilon_2(\mathbf{k}_0)=\lambda_{AC}|f_A(\mathbf{k}_0)|^2$. At the point $\mathbf{k}_0$, the degeneracy of the BTP can now be controlled by the parameters $\lambda_{\alpha\beta}$. If $\lambda_{AB}$ and $\lambda_{AC}$ are nonzero, the flat band is fully gapped. If $\lambda_{AB}=0$ or $\lambda_{AC}=0$, there is a twofold BTP. If $\lambda_{AB}=\lambda_{AC}=0$, there is a threefold BTP, however this possibility should be ignored since the flat band would be trivially decoupled.

(3) If one out of the three components of $\ket{f_{\mathbf{k}_0}}$ vanishes (say $f_C(\mathbf{k}_0)=0$ for concreteness), the degeneracy of the BTP is again tunable by the $\lambda_{\alpha\beta}$. For most values of the $\lambda_{\alpha\beta}$, all bands will be gapped from each other at $\mathbf{k}_0$, but choosing $\lambda_{AB}=0$ leads to $\epsilon_0(\mathbf{k}_0)=\epsilon_2(\mathbf{k}_0)=0$, $\epsilon_1(\mathbf{k}_0)=\lambda_{AC}|f_A(\mathbf{k}_0)|^2+\lambda_{BC}|f_B(\mathbf{k}_0)|^2$, and similarly choosing $\lambda_{AC}|f_A(\mathbf{k}_0)|^2+\lambda_{BC}|f_B(\mathbf{k}_0)|^2=0$ leads to $\epsilon_0(\mathbf{k}_0)=\epsilon_2(\mathbf{k}_0)=0$, $\epsilon_1(\mathbf{k}_0)=\lambda_{AB}(|f_A(\mathbf{k}_0)|^2+|f_B(\mathbf{k}_0)|^2)$. By a proper choice of the $\lambda_{\alpha\beta}$, one can thus obtain a twofold BTP and even a threefold BTP \emph{without} trivially decoupling the flat band.
	
(4) If none of the components of $\ket{f_{\mathbf{k}_0}}$ vanishes, a threefold BTP is impossible, but a twofold BTP can be achieved by $\lambda_{AB}\lambda_{AC}|f_A(\mathbf{k}_0)|^2+\lambda_{AB}\lambda_{BC}|f_B(\mathbf{k}_0)|^2+\lambda_{AC}\lambda_{BC}|f_C(\mathbf{k}_0)|^2=0$.

In summary, the band structure of the Hamiltonian (\ref{3Ham}) can exhibit a gapped flat band, a twofold or a threefold BTP, depending on the interplay between the BCLS $\ket{f_\mathbf{k}}$ and the parameters $\lambda_{\alpha\beta}$.

\section{Examples for 3D flat-band models}
\label{Appcube}

\begin{figure}
	\centering
	\includegraphics[width=\columnwidth]{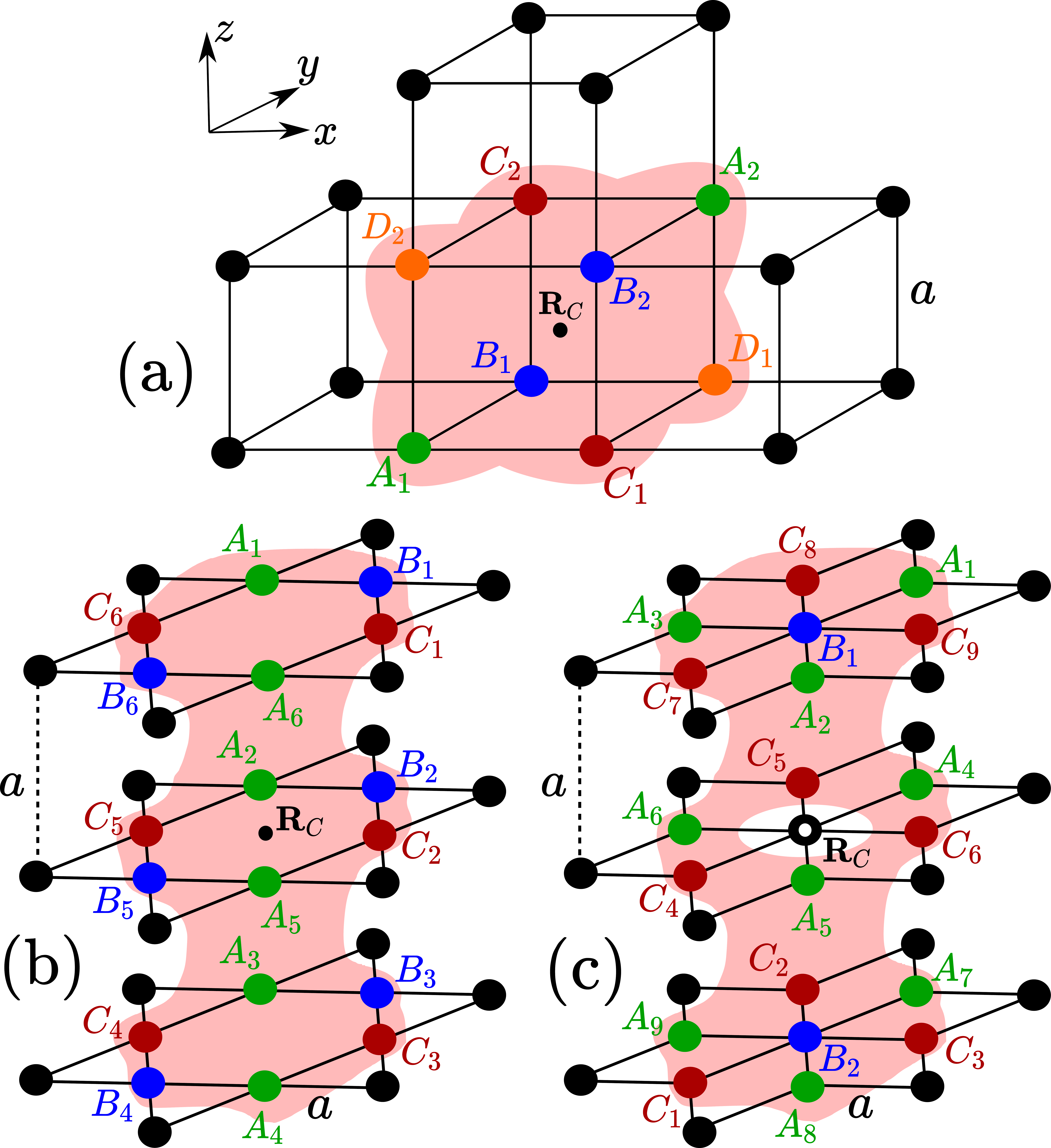}
	\caption{(a) A CLS on an $N=4$ cubic lattice. (b) A CLS on the 3D-stacked Kagome lattice that is type-I CP if the amplitudes are properly correlated. (c) A CLS on the 3D-stacked dice lattice that is type-II CP if the amplitudes are properly correlated.}
	\label{fig:3Dexs}
\end{figure}

Quadratic flat-band models can be built from any CLS on any $d$-dimensional lattice. For example, consider the CLS on the simple 3D cubic lattice shown in Fig. \ref{fig:3Dexs}(a). 
The corresponding BCLS reads
\begin{equation}
\ket{f_\mathbf{k}}=\begin{pmatrix}
w_{A_1}e^{-\frac{i}{2}(k_x+k_y+k_z)}+w_{A_2}e^{\frac{i}{2}(k_x+k_y+k_z)}\\
w_{B_1}e^{-\frac{i}{2}(k_x-k_y+k_z)}+w_{B_2}e^{\frac{i}{2}(k_x-k_y+k_z)}\\
w_{C_1}e^{\frac{i}{2}(k_x-k_y-k_z)}+w_{C_2}e^{-\frac{i}{2}(k_x-k_y-k_z)}\\
w_{D_1}e^{\frac{i}{2}(k_x+k_y-k_z)}+w_{D_2}e^{-\frac{i}{2}(k_x+k_y-k_z)}
\end{pmatrix}.
\end{equation}
Inserting into Eq. (\ref{mainham}), one obtains a flat-band TB model tunable by the CLS amplitudes and six parameters $\lambda_{\alpha\beta}$.

The simplest linear flat-band models on a 3D lattice are three-band models built from chiral or CP-type CLSs. A type-I chiral CLS can be easily found on any lattice. A type-II chiral CLS can be easily constructed for example on a 3D cubic lattice. Finding CP-CLSs is more difficult, but a simple way consists in forming a 3D lattice by stacking up appropriate 2D lattices. Importantly, the localization center must be chosen within one of the 2D layers. For example, a CP$_I$-CLS can be found on the 3D-stacked Kagome lattice, see Fig. \ref{fig:3Dexs}(b), where $w_{\alpha_{i+3}}=-w_{\alpha_i}\in\mathbb{R}$. The corresponding BCLS reads $\ket{f_\mathbf{k}}=|o_\mathbf{k}^{ABC}\rangle=(f_A,f_B,f_C)$, where
\begin{equation}
	\begin{aligned}
	f_A&=-2i[w_{A_1}\text{sin}(k_--k_z)+w_{A_2}\text{sin}k_-\\
	&\hspace{1.1cm}+w_{A_3}\text{sin}(k_-+k_z)],\\
	f_B&=2i[w_{B_1}\text{sin}(k_++k_z)+w_{B_2}\text{sin}k_+\\
	&\hspace{1.1cm}+w_{B_3}\text{sin}(k_+-k_z)],\\
	f_C&=2i[w_{C_1}\text{sin}(k_x+k_z)+w_{C_2}\text{sin}k_x\\
	&\hspace{1.1cm}+w_{C_3}\text{sin}(k_x-k_z)].
\end{aligned}
\end{equation}
Inserting into Eq. (\ref{linim}), one obtains a linear 3D flat-band TB model. Similarly, a CP$_{II}$-CLS can be found on the 3D dice lattice, see Fig. \ref{fig:3Dexs}(c), where $w_{B_1}=w_{B_2}\in\mathbb{R}$ and $w_{C_i}=-w_{A_i}$. The corresponding BCLS reads $\ket{f_\mathbf{k}}=|q_{B,\mathbf{k}}^{ABC}\rangle=(f_A,f_B,-f_A^*)$, where
\begin{equation}
\begin{aligned}
f_A&=w_{A_1}e^{i(k_++k_z)}+w_{A_2}e^{i(k_-+k_z)}+w_{A_3}e^{-i(k_x-k_z)}\\
&+w_{A_4}e^{ik_+}+w_{A_5}e^{ik_-}+w_{A_6}e^{-ik_x}\\
&+w_{A_7}e^{i(k_+-k_z)}+w_{A_8}e^{i(k_--k_z)}+w_{A_9}e^{-i(k_x+k_z)},\\
f_B&=2w_{B_1}\cos k_z.
\end{aligned}
\end{equation}
 Inserting into Eq. (\ref{HABCcon}), one obtains another linear 3D flat-band model.

%
%
%

\section{Construction of linear $N>3$ flat-band Hamiltonians by CLS decomposition}
\label{Nconstr}

As a first step to obtain a linear flat-band Hamiltonian (\ref{mainlin}) from a given CLS on a lattice with $N>3$, decompose the CLS into chiral and/or CP sub-CLSs. This is most conveniently done by identifying in the BCLS (\ref{BCLSN}) all chiral and/or CP sub-BCLSs. 

A \emph{type-I chiral (C$_I$) sub-BCLS} $|m_{\tau,\mathbf{k}}^{\alpha\beta\gamma}\rangle$ is a three-component subvector
\begin{equation}
	\begin{aligned}
		|m_{\alpha,\mathbf{k}}^{\alpha\beta\gamma}\rangle&\equiv(0,f_\beta,f_\gamma)^T,\\
		|m_{\beta,\mathbf{k}}^{\alpha\beta\gamma}\rangle&\equiv(f_\alpha,0,f_\gamma)^T,\\	|m_{\gamma,\mathbf{k}}^{\alpha\beta\gamma}\rangle&\equiv(f_\alpha,f_\beta,0)^T.
		\label{sub2}
	\end{aligned}
\end{equation}
A \emph{type-II chiral (C$_{II}$) sub-BCLS} $|n_{\tau,\mathbf{k}}^{\alpha\beta\gamma}\rangle$ is a three-component subvector
\begin{equation}
	\begin{aligned}
		|n_{\alpha,\mathbf{k}}^{\alpha\beta\gamma}\rangle&=(f_\alpha,-f_\gamma,f_\gamma)^T,\\
		|n_{\beta,\mathbf{k}}^{\alpha\beta\gamma}\rangle&=(f_\alpha,f_\beta,-f_\alpha)^T,\\
		|n_{\gamma,\mathbf{k}}^{\alpha\beta\gamma}\rangle&=(-f_\beta,f_\beta,f_\gamma)^T.
	\end{aligned}
\end{equation}
A \emph{type-I CP (CP$_I$) sub-BCLSs} is a three-component subvector
\begin{equation}
	\begin{aligned}
		|o_\mathbf{k}^{\alpha\beta\gamma}\rangle&\equiv(f_\alpha,f_\beta,f_\gamma)^T,\\
		f_\delta^*&=\kappa_\delta f_\delta,\\ 
		\kappa_\alpha\kappa_\beta\kappa_\gamma&=-1,
	\end{aligned}
	\label{sub1}
\end{equation}
where a sign $\kappa_\delta=1$ ($\kappa_\delta=-1$ ) is assigned to each real (imaginary) component of the BCLS (\ref{BCLSN}). 

A \emph{type-II CP (CP$_{II}$) sub-BCLS}  $|q_{\tau,\mathbf{k}}^{ABC}\rangle$ is a three-component subvector
\begin{equation}
	\begin{aligned}
		|q_{\alpha,\mathbf{k}}^{\alpha\beta\gamma}\rangle&\equiv(f_\alpha,-f_\gamma^*,f_\gamma)^T,\\
		|q_{\beta,\mathbf{k}}^{\alpha\beta\gamma}\rangle&\equiv(f_\alpha,f_\beta,-f_\alpha^*)^T,\\ |q_{\gamma,\mathbf{k}}^{\alpha\beta\gamma}\rangle&\equiv(-f_\beta^*,f_\beta,f_\gamma)^T.
	\end{aligned}
\end{equation}
 If no chiral or CP sub-BCLSs can be found, then the input CLS is not suitable for building a linear flat-band model.

As a second step, for each sub-BCLS found from the above decomposition, one can now construct a matrix that vanishes on $\ket{f_\mathbf{k}}$.
For each type-I chiral sub-BCLS, write down a corresponding $3\times3$ matrix \smash{$\tilde{M}_{\tau,\mathbf{k}}^{\alpha\beta\gamma}$}, where
\begin{equation}
	\begin{aligned}
		\tilde{M}_{\alpha,\mathbf{k}}^{\alpha\beta\gamma}&\equiv\begin{bmatrix}
			0&-f_\gamma&f_\beta\\
			-f_\gamma^*&0&0\\
			f_\beta^* & 0 & 0
		\end{bmatrix}, \\
		\tilde{M}_{\beta,\mathbf{k}}^{\alpha\beta\gamma}&\equiv\begin{bmatrix}
			0&-f_\gamma^*&0\\
			-f_\gamma&0&f_\alpha\\
			0 & f_\alpha^* & 0
		\end{bmatrix}, \\
		\tilde{M}_{\gamma,\mathbf{k}}^{\alpha\beta\gamma}&\equiv\begin{bmatrix}
			0&0&-f_\beta^*\\
			0&0&f_\alpha^*\\
			-f_\beta & f_\alpha & 0
		\end{bmatrix}.
	\end{aligned}
\end{equation} 
Extend each such matrix to an $N\times N$ matrix \smash{$M_{\tau,\mathbf{k}}^{\alpha\beta\gamma}$} by adding zeros in all rows and columns not indicated by $(\alpha,\beta,\gamma)$. Similarly, for each type-II chiral sub-BCLS, write down a corresponding $3\times3$ matrix \smash{$\tilde{N}_{\tau,\mathbf{k}}^{\alpha\beta\gamma}$}, where
\begin{equation}
	\begin{aligned}
		\tilde{N}_{\alpha,\mathbf{k}}^{\alpha\beta\gamma}&\equiv\begin{bmatrix}
			0&f_\gamma^*&f_\gamma^*\\
			f_\gamma&0&-f_\alpha\\
			f_\gamma & f_\alpha & 0
		\end{bmatrix},\\
		\tilde{N}_{\beta,\mathbf{k}}^{\alpha\beta\gamma}&\equiv\begin{bmatrix}
			0 & f_\alpha & f_\beta\\
			f_\alpha^* & 0 & f_\alpha^*\\ 
			-f_\beta & f_\alpha & 0
		\end{bmatrix},\\	\tilde{N}_{\gamma,\mathbf{k}}^{\alpha\beta\gamma}&\equiv\begin{bmatrix}
			0&-f_\gamma&f_\beta\\
			f_\gamma&0&f_\beta\\
			f_\beta^* & f_\beta^* & 0
		\end{bmatrix},
	\end{aligned}
\end{equation}
and extend each such matrix to an $N\times N$ matrix \smash{$N_{\tau,\mathbf{k}}^{\alpha\beta\gamma}$}.
For each type-I CP sub-BCLS, write down a corresponding $3\times3$ matrix
\begin{equation}
	\tilde{O}_\mathbf{k}^{\alpha\beta\gamma}\equiv\begin{bmatrix}
		0&-f_\gamma&f_\beta\\
		-\mu_\gamma f_\gamma&0&\mu_\gamma f_\alpha\\
		\mu_\beta f_\beta & -\mu_\beta f_\alpha & 0
	\end{bmatrix},
\end{equation} 
and extend each such matrix to an $N\times N$ matrix \smash{$O_\mathbf{k}^{\alpha\beta\gamma}$}. Finally, for each type-II CP sub-BCLS, write down a corresponding $3\times3$ matrix \smash{$\tilde{Q}_{\tau,\mathbf{k}}^{\alpha\beta\gamma}$}, where
\begin{equation}
	\begin{aligned}
		\tilde{Q}_{\alpha,\mathbf{k}}^{\alpha\beta\gamma}&\equiv\begin{bmatrix}
			0&f_\gamma&f_\gamma^*\\
			f_\gamma^*&f_\alpha&0\\
			f_\gamma & 0 & -f_\alpha
		\end{bmatrix},\\
		\tilde{Q}_{\beta,\mathbf{k}}^{\alpha\beta\gamma}&\equiv\begin{bmatrix}
			-f_\beta & f_\alpha & 0\\
			f_\alpha^* & 0 & f_\alpha\\
			0 & f_\alpha^* & f_\beta
		\end{bmatrix},\\	\tilde{Q}_{\gamma,\mathbf{k}}^{\alpha\beta\gamma}&\equiv\begin{bmatrix}
			f_\gamma&0&f_\beta^*\\
			0&-f_\gamma&f_\beta\\
			f_\beta & f_\beta^* & 0
		\end{bmatrix},
	\end{aligned}
\end{equation}
and extend each such matrix to an $N\times N$ matrix \smash{$Q_{\tau,\mathbf{k}}^{\alpha\beta\gamma}$}.

The idea is now to form a linear combination out of all the $N\times N$ matrices found from this procedure. However, for getting a reasonable Bloch Hamiltonian, we have to recall the discussion of Section \ref{lin3exs}: not all chiral or CP-type CLSs lead to linear $N=3$ Hamiltonians that make sense on a lattice. Similarly, among all the sub-CLSs into which a given CLS on a lattice with $N$ orbitals can be decomposed, not all will lead to matrices \smash{$M_{\tau,\mathbf{k}}^{\alpha\beta\gamma}$}, \smash{$N_{\tau,\mathbf{k}}^{\alpha\beta\gamma}$}, \smash{$O_\mathbf{k}^{\alpha\beta\gamma}$} and \smash{$Q_{\tau,\mathbf{k}}^{\alpha\beta\gamma}$} that are physically reasonable. We will thus only retain physically reasonable matrices.

As a last step to obtain a linear flat-band Hamiltonian, form a linear combination 
 \begin{equation}
	H_\mathbf{k}=\sum_{\alpha,\beta>\alpha,\gamma>\beta}\lambda_\mathbf{k}^{\alpha\beta\gamma}F_\mathbf{k}^{\alpha\beta\gamma}
\end{equation} 
out of all the (physically reasonable) matrices found from the above procedure, where \smash{$F_\mathbf{k}^{\alpha\beta\gamma}=M_{\tau,\mathbf{k}}^{\alpha\beta\gamma}$}, \smash{$F_\mathbf{k}^{\alpha\beta\gamma}=N_{\tau,\mathbf{k}}^{\alpha\beta\gamma}$}, \smash{$F_\mathbf{k}^{\alpha\beta\gamma}=O_\mathbf{k}^{\alpha\beta\gamma}$}, or \smash{$F_\mathbf{k}^{\alpha\beta\gamma}=Q_{\tau,\mathbf{k}}^{\alpha\beta\gamma}$}. This is Eq. (\ref{mainlin}) provided in the main text.

As an example for this procedure, consider the CLS shown in Fig. \ref{fig:conjchiral}(a), with corresponding BCLS (\ref{BCLSex5}). In this BCLS, we can identify six type-I chiral sub-BCLSs, namely $|m_{D,\mathbf{k}}^{ABD}\rangle$, $|m_{D,\mathbf{k}}^{ACD}\rangle$, $|m_{D,\mathbf{k}}^{ADE}\rangle$, $|m_{D,\mathbf{k}}^{BCD}\rangle$, $|m_{D,\mathbf{k}}^{BDE}\rangle$, and $|m_{D,\mathbf{k}}^{CDE}\rangle$. Moreover, we can find one type-I CP sub-BCLS, namely $|o_\mathbf{k}^{ABC}\rangle$. For each of these sub-BCLSs, we may now construct an $N\times N$ matrix that vanishes on the BCLS (\ref{BCLSex5}). For instance, from $|m_{D,\mathbf{k}}^{ABD}\rangle$ and $|m_{D,\mathbf{k}}^{ACD}\rangle$, we get
\[
M_{D,\mathbf{k}}^{ABD}=\begin{bmatrix}
0 & 0 & 0 & -f_B^* & 0\\
0 & 0 & 0 & f_A^* & 0\\
0 & 0 & 0 & 0 & 0\\
-f_B & f_A & 0 & 0 & 0\\
0 & 0 & 0 & 0 & 0
\end{bmatrix},\]
and
\[
M_{D,\mathbf{k}}^{ACD}=\begin{bmatrix}
	0 & 0 & 0 & -f_C^* & 0\\
	0 & 0 & 0 & 0 & 0\\
	0 & 0 & 0 & f_A^* & 0\\
	-f_C & 0 & f_A & 0 & 0\\
	0 & 0 & 0 & 0 & 0
\end{bmatrix},\]
respectively.
The former matrix is not physically reasonable, as can be seen in the following way: if it contributed to $H_\mathbf{k}$, then in the real-space tight-binding model there would have to be a hopping from $A$ to $D$ orbitals along directions $\pm\frac{1}{2}(\mathbf{e}_x+\sqrt{3}\mathbf{e}_y)$ and $\pm\frac{1}{2}(3\mathbf{e}_x-\sqrt{3}\mathbf{e}_y)$. Clearly, such hoppings are not compatible with the lattice, cf. Fig. \ref{fig:conjchiral}(a). Thus, the sub-CLS $\ket{\psi_{ABD}}$ is not physically reasonable. In contrast, the hoppings arising when $M_{D,\mathbf{k}}^{ACD}$ contributes to $H_\mathbf{k}$ make sense on the lattice. Thus, the sub-CLS $\ket{\psi_{ACD}}$ is physically reasonable. Proceeding in the same way, one finds that there are three reasonable sub-CLSs, shown in  Fig. \ref{fig:conjchiral}(b), and therefore the Hamiltonian (\ref{N5conj}) is a linear combination of the three corresponding matrices.

Similarly, for the CLS of Fig. \ref{fig:spinchiral}(a), with corresponding BCLS (\ref{BCLSexx5}), we can identify six type-I chiral sub-BCLSs, namely $|m_{E,\mathbf{k}}^{ABE}\rangle$, $|m_{E,\mathbf{k}}^{ACE}\rangle$, $|m_{E,\mathbf{k}}^{ADE}\rangle$, $|m_{E,\mathbf{k}}^{BCE}\rangle$, $|m_{E,\mathbf{k}}^{BDE}\rangle$ and $|m_{E,\mathbf{k}}^{CDE}\rangle$, and one type-II CP sub-BCLS $|q_{B,\mathbf{k}}^{ABC}\rangle$. Among these, there are three reasonable sub-CLSs, shown in Fig. \ref{fig:spinchiral}(b), and the linear Hamiltonian (\ref{N5spin}) can be built from the corresponding matrices.

\vspace{.5cm}

\section{Nearest-neighbor model on the Kagome lattice from a BCLS perspective}
\label{Appkagopyro}

The nearest-neighbor TB model on the two-dimensional $N=3$ Kagome lattice [cf. Fig. \ref{fig:newCLSexample}(b) for conventions] with uniform hopping amplitudes $t=1/2$  is captured by a Bloch Hamiltonian of the form 
\begin{equation}
H_\mathbf{k}=\begin{bmatrix}
0 & \cos k_x & \cos k_+\\
\cos k_x & 0 & \cos k_-\\
\cos k_+ & \cos k_- & 0
\end{bmatrix},
\end{equation}
and the unnormalized flat-band eigenstate (BCLS) reads
\begin{equation}
\ket{f_\mathbf{k}}=(f_A,f_B,f_C)^T=(\sin k_-, \sin k_+, - \sin k_x)^T.
\label{kagofk}
\end{equation}
Clearly, the Bloch Hamiltonian can thus be written (upon a proper sign choice) as
\begin{equation}
	H_\mathbf{k}=\begin{bmatrix}
		0 & \sqrt{1-f_C^2} & \sqrt{1-f_B^2}\\
		\sqrt{1-f_C^2} & 0 & \sqrt{1-f_A^2}\\
		\sqrt{1-f_B^2} &  \sqrt{1-f_A^2} & 0
	\end{bmatrix},
\label{squareroot}
\end{equation}
i.e. the matrix elements are infinite-order polynomials in the BCLS components, in contrast to the quadratic-order Hamiltonian (\ref{mainham}) and the linear-order Hamiltonian (\ref{mainlin}) introduced in this paper. Importantly, however, the Hamiltonian (\ref{squareroot}) does \emph{not} represent a generic class of flat-band models: the flat-band property $H_\mathbf{k}\ket{f_\mathbf{k}}=\epsilon_0\ket{f_\mathbf{k}}$ (with $\epsilon_0=-1$) is verified only due to the very specific symmetries of the Kagome BCLS (\ref{kagofk}), but is absent for generic $\ket{f_\mathbf{k}}=(f_A,f_B,f_C)^T$.

\bibliography{ref}

\end{document}